\documentclass[11pt,a4paper]{emulateapj}
\usepackage{graphicx}        
\usepackage{color}           
\usepackage{url}             

\usepackage{epsfig}
\usepackage{amsmath}

\newcommand{\beq}{\begin{equation}}
\newcommand{\eeq}{\end{equation}}
\newcommand{\beqa}{\begin{eqnarray}}
\newcommand{\eeqa}{\end{eqnarray}}

\newcommand{\grad}{ {\bf \nabla } }

\begin{document}



\title{Numerical simulations of impulsively generated Alfv\'en waves in solar magnetic arcades}

\author{P.~Chmielewski$^{1}$,
        K.~Murawski$^{1}$,
        Z.E.~Musielak$^{2,3}$,
	A.K.~Srivastava$^{4}$     
       }
\shortauthors{P.~Chmielewski et al.}
\shorttitle{Alfv\'en waves in solar magnetic arcades}

\altaffiltext{1}{Group of Astrophysics, UMCS, ul. Radziszewskiego 10, 20-031 Lublin, Poland}
\altaffiltext{2}{Department of Physics, University of Texas at Arlington, Arlington, TX 76019, USA}
\altaffiltext{3}{Kiepenheuer-Institut f\"ur Sonnenphysik, Sch\"oneckstr. 6, 79104 Freiburg, Germany}
\altaffiltext{4}{Department of Physics, Indian Institute of Technology (Banaras Hindu University), Varanasi-221005, India}

\begin{abstract}
We perform numerical simulations of impulsively generated Alfv\'en waves in 
an isolated solar arcade, which is gravitationally stratified and magnetically 
confined.  We study numerically the propagation of Alfv\'en waves {along 
such magnetic structure that extends from the lower chromosphere, where the 
waves are generated, to the solar corona, and analyze influence of the arcade 
size and width of the initial pulses on the wave propagation and reflection.}  
Our model of the solar atmosphere is constructed by adopting the temperature 
distribution based on the semi-empirical VAL-C model and specifying the curved 
magnetic field lines that constitute the asymmetric magnetic arcade.  The 
propagation and {reflection} of Alfv\'en waves in this arcade is described 
by 2.5D magnetohydrodynamic equations that are numerically solved by the FLASH 
code.  Our numerical simulations reveal that the Alfv\'en wave amplitude decreases 
as a result of a partial reflection of Alfv\'en waves in the solar transition 
region, and {that the waves which are not reflected leak through the 
transition region and reach the solar corona.}  We also find the decrement of 
the attenuation time of Alfv\'en waves for wider initial pulses.  Moreover, our 
results show that the propagation of Alfv\'en waves in the arcade is affected 
by spatial dependence of the Alfv\'en speed, which leads to phase-mixing that 
is stronger for more curved and larger magnetic arcades.  We discuss processes 
that affect the Alfv\'en wave propagation {in an asymmetric solar arcade} 
and conclude that besides phase-mixing in the magnetic field configuration, 
plasma properties of the arcade and size of the initial pulse as well as 
structure of the solar transition region all play a vital role in the Alfv\'en 
wave propagation. 
\end{abstract}

\keywords{Sun: atmosphere, Sun : corona, magnetohydrodynamics (MHD), magnetic fields}

%
\section{Introduction}
%
Arcades are gravitationally stratified and magnetically confined structures 
in the solar atmosphere.  Their physical properties are determined by flows of 
plasma along these arcades 
and the heating resulting from the energy dissipation by 
different waves (\v Cadez et al. 1994, Innes et al. 2003, McKenzie \& Savage 2009).  
Investigations of magnetic topology of such arcades (Biskamp \& Welter 1989, 
Miki\'c et al. 1989) and magnetohydrodynamic (MHD) wave propagation in these 
structures (e.g., Del Zanna et al. 2005, Selwa et al. 2005, Selwa et al. 
2006, D\'iaz et al. 2006) are potential areas of research that reveal both 
the wave heating and plasma dynamics.  In more recent studies carried on by 
Gruszecki \& Nakariakov (2011), the excitation of slow magnetoacustic waves 
was investigated in a system of magnetic arcades.  Specifically, the 
antisymmetric kink mode of magnetoacustic waves was modelled in the magnetic 
arcades and used for determination of the arcade magnetic field, which is often 
referred as a MHD seismology.

Theoretical predictions resulting from the above studies have been supplemented 
by several significant efforts to observe such waves in the magnetic arcades.
Verwichte et al. (2004) detected the damped standing kink oscillations in the 
arcades using the Transition Region and Coronal Explorer (TRACE) observations. 
Verwichte et al. (2010) reported on the transversal kink oscillations in the 
coronal arcades obtained by SOHO/EIT and STEREO observations, and used the 
results to invoke the spatial MHD seismology of these magnetic structures
to estimate their local plasma conditions.

The previous studies of magnetoacoustic waves were also supplemented by 
investigations of the propagation of purely incompressible transverse 
(Alfv\'en) waves in such magnetic arcades in the solar atmosphere.  Since 
the magnetic field serves as a guide for Alfv\'en waves, it is worth 
to study these waves in magnetic field structures like arcades. Arregui et 
al. (2004) studied MHD waves, including Alfv\'en, coupled Alfv\'en and 
fast magnetoacoustic (fast, henceforth) waves, and the magnetic arcade 
oscillations connected with the propagation 
of these waves.  In a cold plasma approximation, they showed that the linear 
coupling of discrete fast modes, which are characterized by a discrete spectrum 
of frequencies and a global velocity structure, and Alfv\'en continuum modes,
which are characterized by a continuous spectrum of frequencies and a velocity 
perturbation confined to given magnetic surfaces, leads to modes with mixed 
properties that arise in the magnetic arcades. 

Similarly, Oliver et al. (1996) also investigated the mixed fast magnetoacoustic 
and Alfv\'en waves in the form of MHD perturbations excited in the magnetic arcades.  
Due to the resonant coupling, Rial et al. (2010) found that the 
fast wave transfers its energy to Alfv\'enic oscillations 
localized around a particular magnetic surface within the arcade, thus producing 
the attenuation of the initial fast MHD mode.  Moreover, they discussed another case 
in which the generated fast wavefront leaves its energy on several magnetic 
surfaces within the arcade.  The system is therefore able to trap energy in the 
form of Alfv\'enic oscillations, which are subsequently phase-mixed to smaller 
spatial scales.
Gruszecki et al. (2007)
considered the energy loss by Alfv\'en waves for straight and curved magnetic 
fields, and found a relation between the energy loss and the amplitude of the initial
wave pulse. Results clearly showed that Alfv\'en waves in a magnetic arcade 
experience a decrease of their amplitude and that this decrease is caused by 
partial reflection of the waves from the solar transition region (Gruszecki at al. 
2007). The partial reflection of Alfv\'en waves results from a steep gradient of 
Alfv\'en speed in the transition region, because of a significant density drop 
in this region of the solar atmosphere. 

{
An important problem of Alfv\'en waves chromospheric leakage in coronal loops 
was studied by Ofman~(2002) in a normalized visco-resistive nonlinear 1.5D 
MHD model, for which the Alfv\'en leakage time was computed.  Moreover, Ofman \& 
Wang~(2007) constructed a MHD model that allowed them to examine the periodic 
oscillations in their coronal loop and compare the results those observed by 
Solar Optical Telescope (SOT) on the board of Hinode.}

Del Zanna et al. (2005) performed numerical studies of {Alfv\'en waves 
triggered by solar flare events} in an isothermal and symmetric arcade model, 
and showed that these waves propagate back and forth along the arcade and 
rapidly decay.  According to the authors, the fast decay of these waves 
explains the observed damping of the amplitudes of oscillations, and the 
efficiency of Alfv\'en waves decay is related to the stratification of solar 
atmosphere as well as to a gradient of magnetic field along the arcade.
{
Miyagoshi et al.~(2004) investigated oscillations of a symmetric solar arcade
and show that the amplitude of the loop oscillations decreases exponentially 
in time due to the energy transport by fast-mode MHD waves.
}

Based on the above results, we may conclude that curved magnetic fields, such as 
those existing in solar magnetic arcades, effectively increase the wave energy 
leakage.  Moreover, the stronger Alfv\'en wave pulse amplitudes are the more efficient 
generation of fast magnetoacoustic waves occurs.  Nevertheless, additional studies, 
in which {the solar atmosphere model is extended to include the solar chromosphere 
and transition region and a broader range of physical parameters for an asymmetric 
magnetic field configuration is considered}, are urgently needed.  Therefore, the 
main goal of this paper is to perform such studies and significantly extend the 
previous work of Del Zanna et al. (2005) and Miyagoshi et al.~(2004).

In our approach, we consider an asymmetric, stratified and non-isothermal solar 
arcade, assume that the temperature profile along the arcade is based on the 
VAL-C semi-empirical model of the solar atmosphere (Vernazza et al. 1981), and 
follow Low (1985) to construct a magnetic field model of the arcade.  An 
important difference between our approach and that presented by Del Zanna et al. 
{and Miyagoshi et al.} is that we impulsively generate Alfv\'en waves by 
launching them in the solar chromosphere, just above the solar photosphere,
whereas in {Del Zanna's} work the waves are triggered by an onset of solar 
flare in the solar corona.  {Since flares are highly episodic transients 
that only occur in solar active regions, where even more complex loops can exist 
with highly changeable ambient medium, Del Zanna et al. (2005) model is designed 
to study short lived Alfv\'en waves in post flare loops, which are most dynamic 
and transient loop systems.  Similarly, Miyagoshi et al. (2004) implemented 
transversal velocity pulses at the top of the coronal loop to excite the waves.  
In contrary, our model allows us to investigate the effects caused by trains of 
Alfv\'en waves resulting from wave reflection in the solar transition region in 
a stable and more realistic solar coronal arcade.  As a consequence of this 
fundamental difference between these two models, we are able to explore 
different physical aspects of the Alfv\'en wave propagation in the arcade 
than those studied by Del Zanna et al. (2005) and Miyagoshi et al. (2004).
}

In this paper, we use the publicly available numerical code FLASH to 
perform parametric studies of the propagation of impulsively generated Alfv\'en 
waves in the arcade by varying an initial pulse position and its width.  Our 
model of solar arcade and the location of Alfv\'en wave sources in the solar 
chromosphere (just above the solar photosphere) allows us to investigate the 
efficiency of Alfv\'en wave propagation through the solar transition region, 
the resulting partial wave reflection, and the wave energy leakage through 
this region.  We use our numerical results to identify the basic physical 
processes that affect the propagation of Alfv\'en waves in our solar arcade 
model.

The paper is organized as follows. Our model of solar magnetic arcades and 
description of our numerical method are introduced in Sects.~2 and~3,
respectively.  Results of our numerical simulations of the Alfv\'en wave 
propagation in magnetic arcades are presented and discussed in Sect. 4.
Conclusions are given in Sect. 5.
%
\section{Numerical  model of Alfv\'en waves}\label{sec:atm_model}
%
%
\subsection{MHD equations}\label{sec:equ_model}
%
We consider a gravitationally stratified and magnetically confined plasma in a 
structure that resembles an arcade, which is described by the following set of 
ideal MHD equations:
%
\beqa
\label{eq:MHD_rho} 
{{\partial \varrho}\over {\partial t}}+\nabla \cdot (\varrho{\bf V})=0\, ,\\
\label{eq:MHD_V}
\varrho{{\partial {\bf V}}\over {\partial t}}+ \varrho\left ({\bf V}\cdot \nabla\right )
{\bf V}= -\nabla p+ \frac{1}{\mu} (\nabla\times{\bf B})\times{\bf B} +\varrho{\bf g}\, , \\
\label{eq:MHD_B}
{{\partial {\bf B}}\over {\partial t}}= \nabla \times ({\bf V}\times {\bf B})\, , \\
\label{eq:MHD_divB}
\nabla\cdot{\bf B} = 0\, , \\
\label{eq:MHD_p}
{\partial p\over \partial t} + {\bf V}\cdot\nabla p = -\gamma p \nabla \cdot {\bf V}\, ,\\
\label{eq:MHD_CLAP}
p = \frac{k_{\rm B}}{m} \varrho T\, ,
\eeqa
%
where ${\varrho}$ is mass density, $p$ is gas pressure, ${\bf V}$, ${\bf B}$ and ${\bf g}
=(0,-g,0)$ represent the plasma velocity, the magnetic field and gravitational 
acceleration, respectively.  In addition, $T$ is a temperature, $m$ is a particle mass,
that was specified by mean molecular weight value of $1.24$ (Oskar Steiner, private communication), 
$k_{\rm B}$ is Boltzmann's constant, $\gamma=5/3$ is the adiabatic index, 
and $\mu$ is the magnetic permeability of plasma. 
The value of $g$ is equal to $274$ m s$^{-2}$.

Note that Del Zanna et al. (2005) worked in the framework of isothermal plasma assumption, $\gamma=1$, 
which resulted in the elimination of energy equation 
 and 
 thermal pressure was described by the ideal gas law of Eq.~(\ref{eq:MHD_CLAP}). 
 This approximation 
greatly simplifies numerical methods, 
 particularly in regions of 
 strongly magnetized plasma, 
 where negative gas pressure can set in. 

%
\begin{figure*}
	\begin{center}
		\includegraphics[width=10.0cm, angle=0]{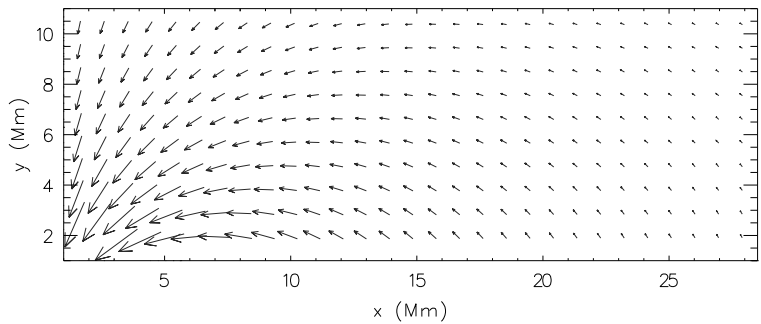}
		\caption{\small Vectors of equilibrium magnetic field.} 
		\label{fig:mag}
	\end{center}
\end{figure*}
%

%
\begin{figure}
	\begin{center}
		\includegraphics[width=7.2cm]{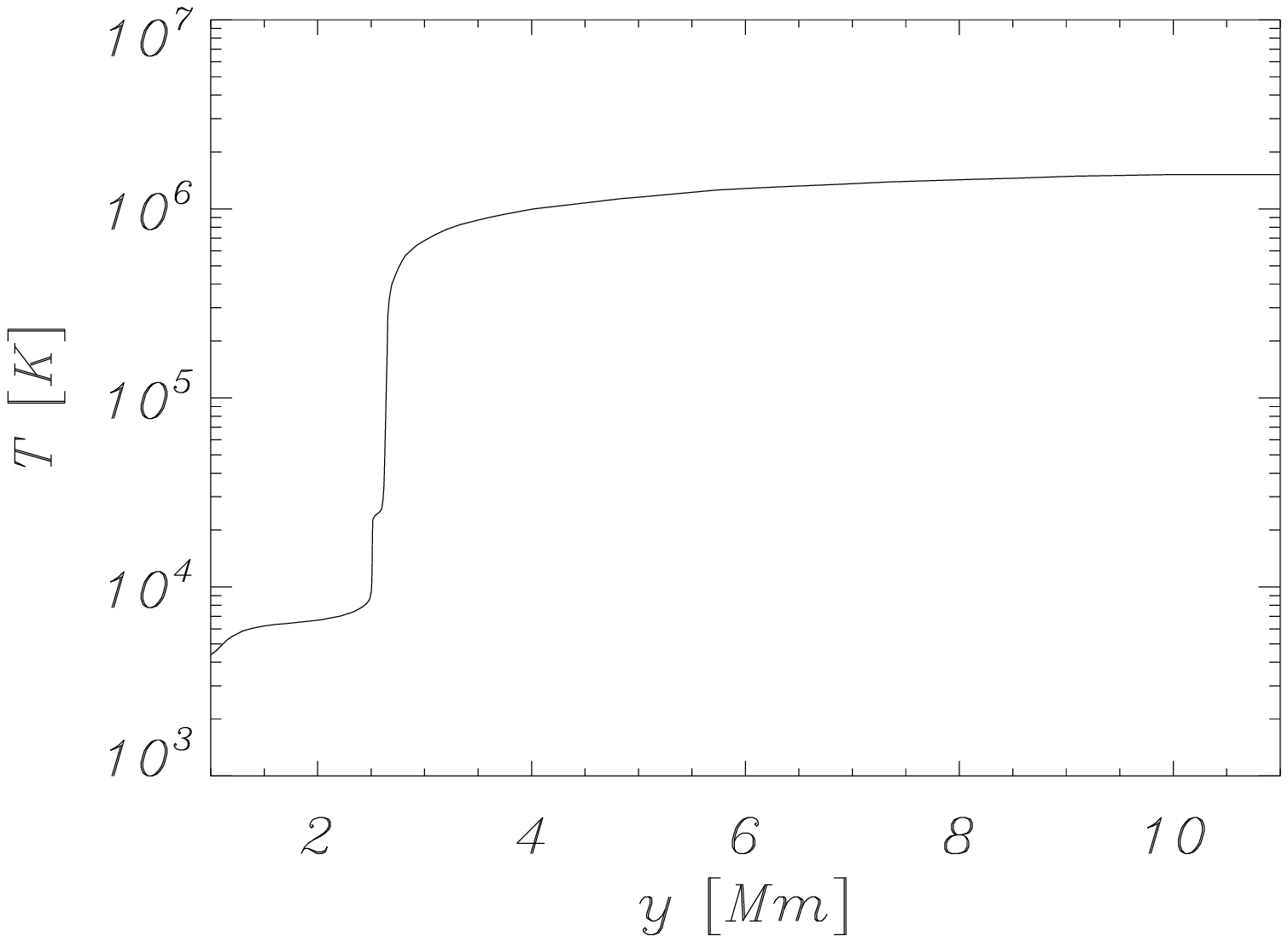}
		\includegraphics[width=7.2cm]{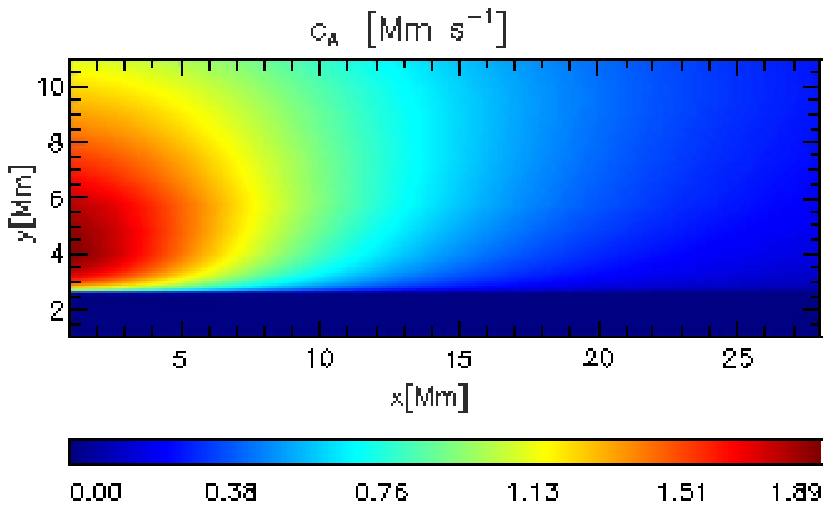}
		\caption{\small Equilibrium profiles of the temperature along a fixed $x$ (top panel)
			and the Alfv\'en speed (bottom panel). 
			} 
		\label{fig:TeAlf}
	\end{center}
\end{figure}
%


%
%
\begin{figure*}
	\begin{center}
		\includegraphics[width=11.2cm, angle=180]{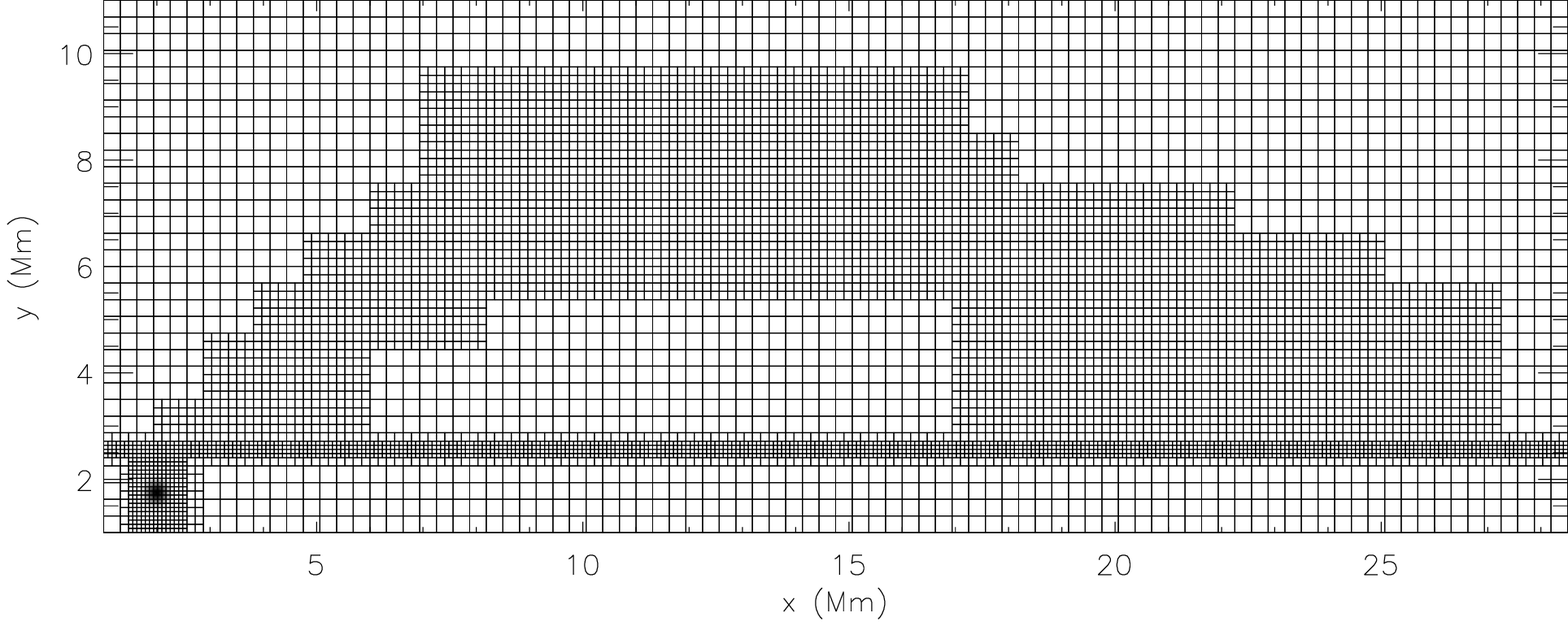}
		\caption{\small Numerical blocks used in the numerical simulations for the case of $x_{\rm 0}=3$~Mm. }
		\label{fig:blk}
	\end{center}
\end{figure*}
%

%
\subsection {A model of the static solar atmosphere}\label{sec:equil}
%

%
We consider a model of the static ($\partial/\partial t = 0$) solar atmosphere with 
an invariant coordinate $z$ 
($\partial/\partial z = 0$), but allow the $z$-components of velocity ($V_{\rm z}$) 
and magnetic field ($B_{\rm z}$) to vary with $x$ and $y$. 
In such 2.5 dimensional (2.5D) 
model, the solar atmosphere is in static equilibrium (${\bf V}_{\rm e}={\bf 0}$) 
with the force-free and current-free magnetic field defined by
%
\beq\
(\nabla\times{\bf B}_{\rm e})\times{\bf B}_{\rm e} = {\bf 0}\, , \hspace{4mm} 
\nabla\times {\bf B}_{\rm e}={\bf 0}\, .
\label{eq:B}
\eeq
%
Here the subscript $_{\rm e}$ corresponds to equilibrium quantities.
We adopt a realistic magnetic flux-tube model originally developed 
for 3D by Low (1985), in which
%
\beq\
{\bf B}_{\rm e}(x,y) = \nabla\times \left( A {\bf\hat{z}} \right) \, , 
\label{eq:Be}
\eeq
%
is an equilibrium magnetic field
with ${\bf\hat{z}}$ being a unit vector along the $z$-direction and $A$ denoting 
the magnetic flux function given by
%
\beq\
A(x,y) = \frac {x (y_{\rm ref} - b)^2 } {(y-b)^2 - x^2} B_{\rm ref}  \, . 
\label{eq:Be2}
\eeq
%
Here $b$ is a constant, which we choose and set equal to, $b=-5$~Mm, and $B_{\rm ref}$ 
is the magnetic field at the reference level, $y_{\rm ref}$ = 10 Mm.
Vectors of magnetic field, resulting from Eq.~(\ref{eq:Be}), are displayed in Fig.~\ref{fig:mag}.
The magnetic field is defined by the magnetic pole that is located at the point ($x$ = 
0 Mm, $y$ = $-$5 Mm).  Note that at a given altitude $y$ magnetic field is strongest around 
the line $x=0$ Mm, where it is essentially vertical.  However, further out the magnetic 
field declines with larger values of $|x|$, revealing its curved structure.  Such magnetic 
field corresponds to an isolated asymmetric magnetic arcade. 

%
As a result of Eqs.~(\ref{eq:MHD_V}) and~(\ref{eq:B}), the pressure gradient is balanced 
by the force of gravity,
%
\begin{equation}\label{eq:p}
-\nabla p_{\rm e} + \varrho_{\rm e} {\bf g} = {\bf 0}\, .
\end{equation}
%
With the use of the ideal gas law given by Eq. (\ref{eq:MHD_CLAP}) and the $y$-component 
of the hydrostatic pressure balance described by Eq. (\ref{eq:p}), we express the 
equilibrium gas pressure and mass density as
%
\beqa\label{eq:pres} 
p_{\rm e}(y)=p_{\rm ref}~{\rm exp}\left( -
\int_{y_{\rm r}}^{y}\frac{dy^{'}}{\Lambda (y^{'})} \right)\, ,\\
\varrho_{\rm e} (y)=\frac{p_{\rm e}(y)}{g \Lambda (y)}\, ,
\eeqa
%
where
%
\begin{equation} 
\Lambda(y) = \frac{k_{\rm B} T_{\rm e}(y)} {mg}\ ,
\end{equation}
%
is the pressure scale-height, and $p_{\rm ref}$ denotes the gas pressure at the 
reference level.

We adopt a realistic plasma temperature profile given by the semi-empirical VAL-C 
model (Vernazza et~al.~1981) that is extrapolated into the solar corona, 
(Fig.~\ref{fig:TeAlf}, top panel). 
In our model, the temperature attains a value of about 
$5 \times 10^{3}$~K at $y=1.5$ Mm and it increases to about $1.5 \times 10^{6}$ K in 
the solar corona  at $y=10$ Mm.  Higher up in the solar corona the temperature is 
assumed to be constant.  The temperature profile determines uniquely the equilibrium 
mass density and gas pressure profiles.  At the transition region, which is located at 
$y \simeq 2.7$ Mm, $T_{\rm e}$ exhibits an abrupt jump (Fig.~\ref{fig:TeAlf}, top panel),
but $\varrho_{\rm e}(y)$ and $p_{\rm e}(y)$ experience a sudden fall off with the 
atmospheric height (not shown).
%

%
\begin{figure*}
\centering{
           \includegraphics[width=4.2cm,angle=0]{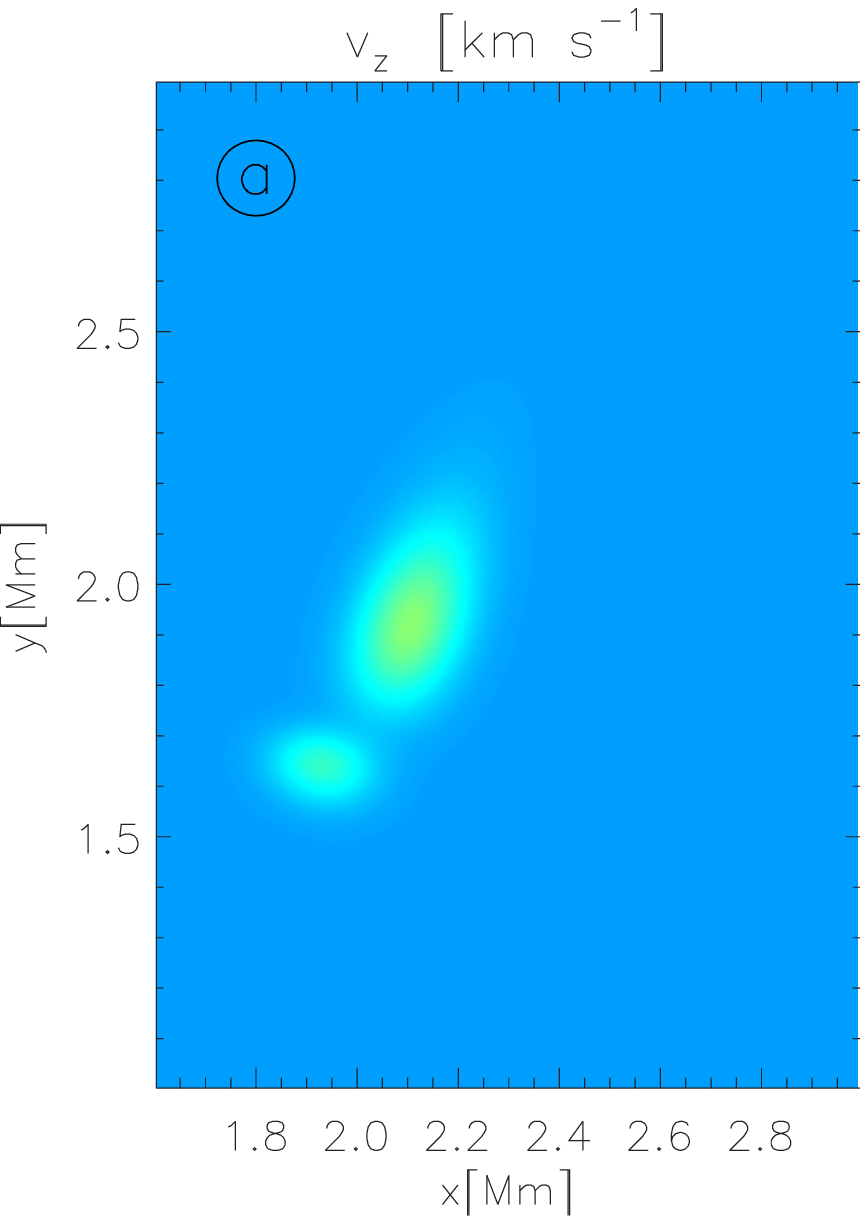}
           \includegraphics[width=4.2cm,angle=0]{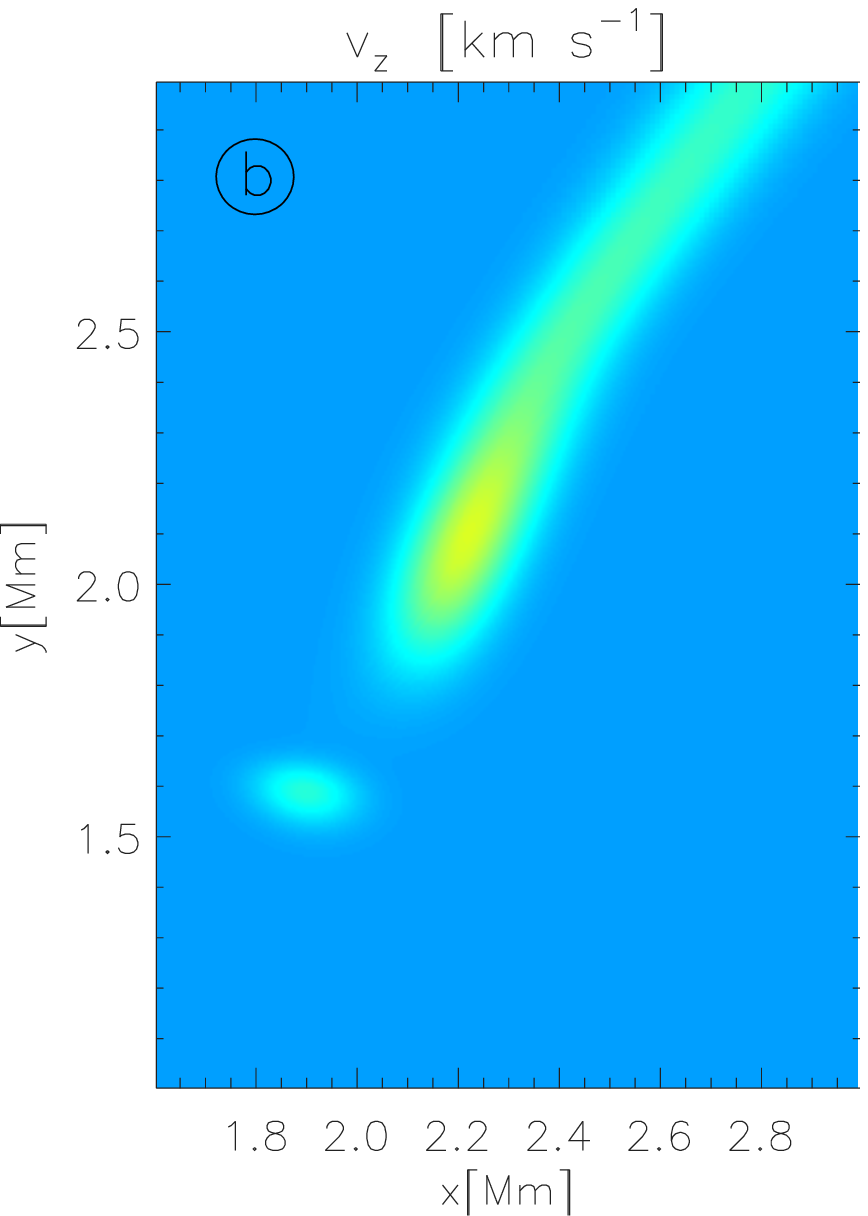}
           \includegraphics[width=4.2cm,angle=0]{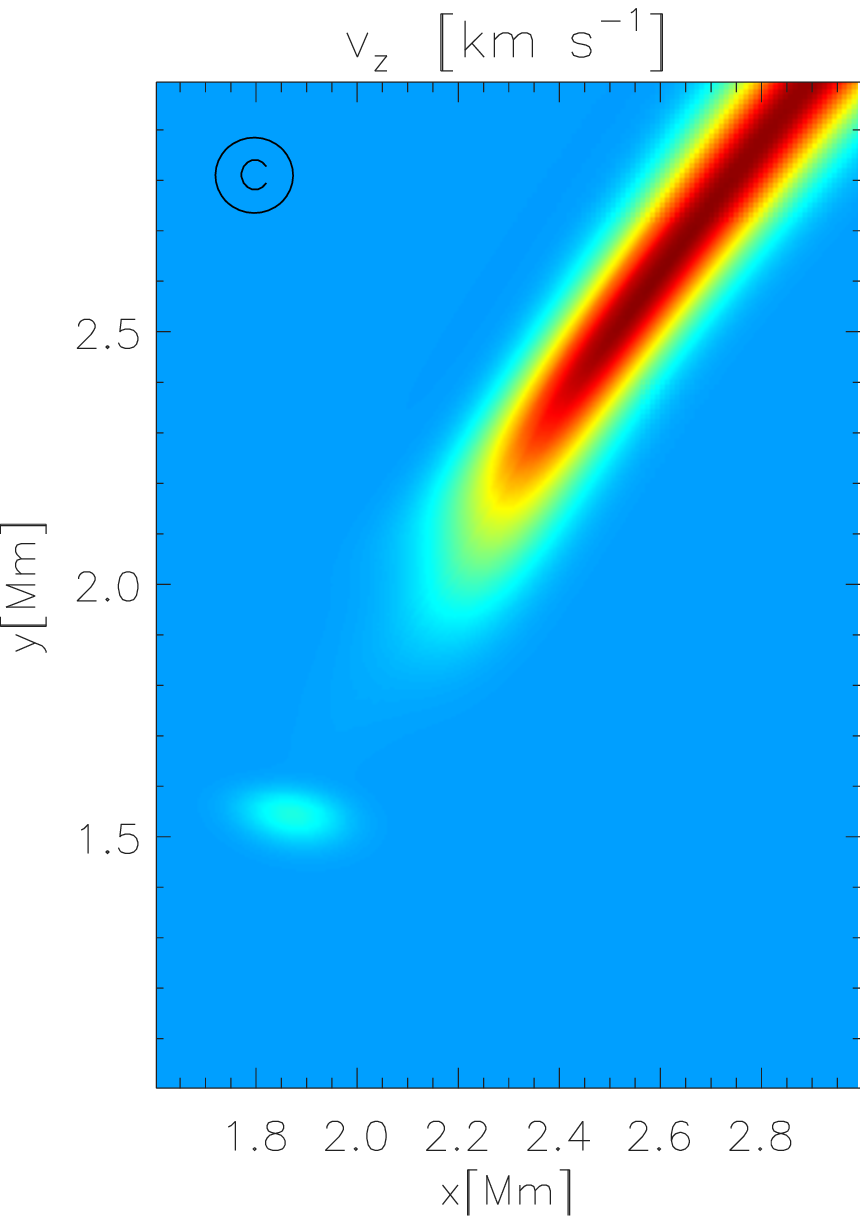}\\\vspace{0.2cm}
           \includegraphics[width=4.2cm,angle=0]{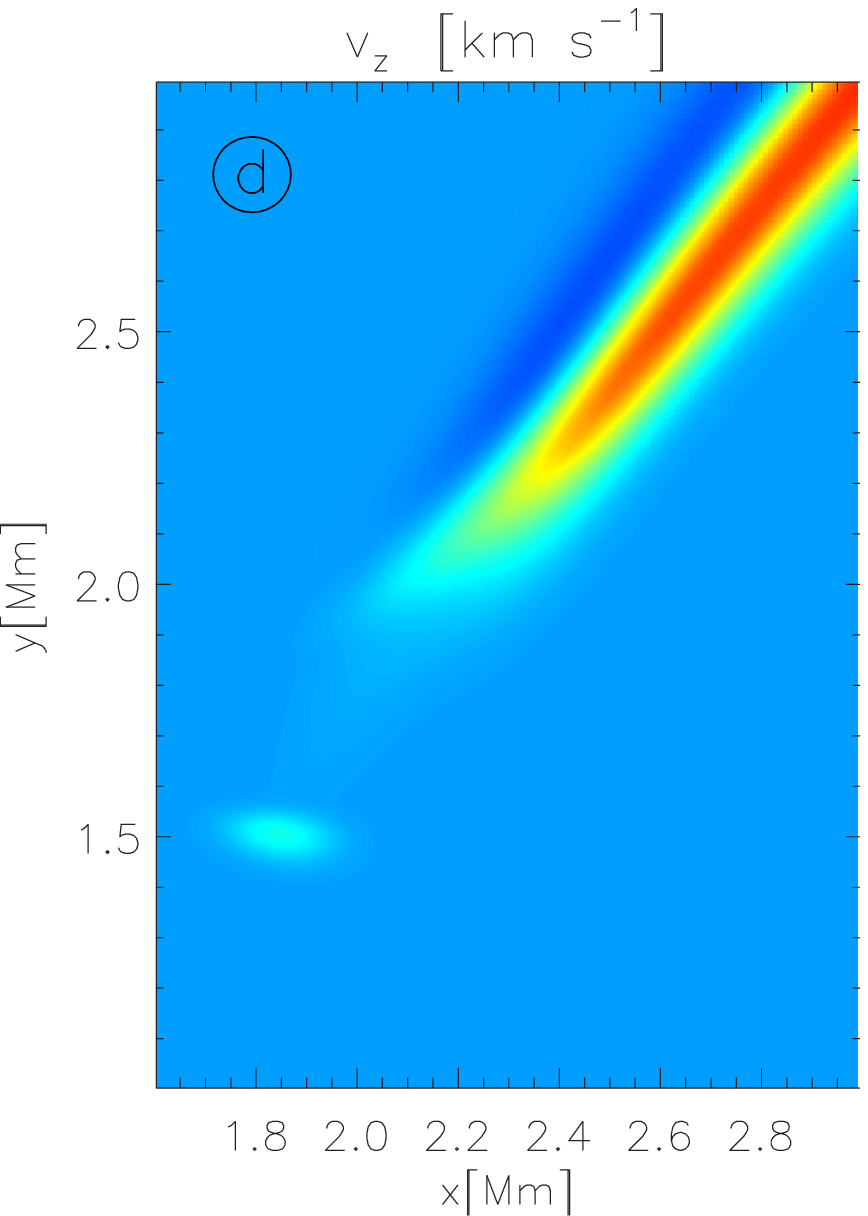}
           \includegraphics[width=4.2cm,angle=0]{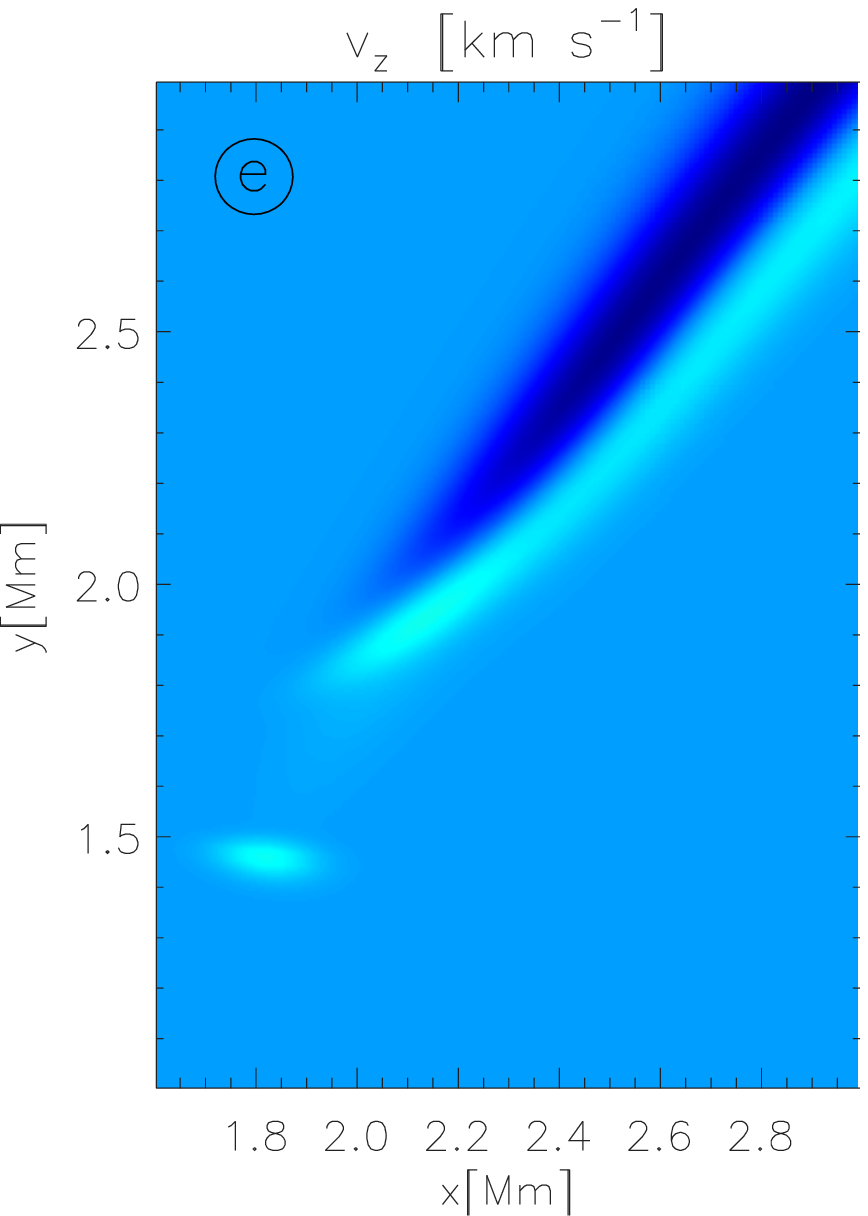}
           \includegraphics[width=4.2cm,angle=0]{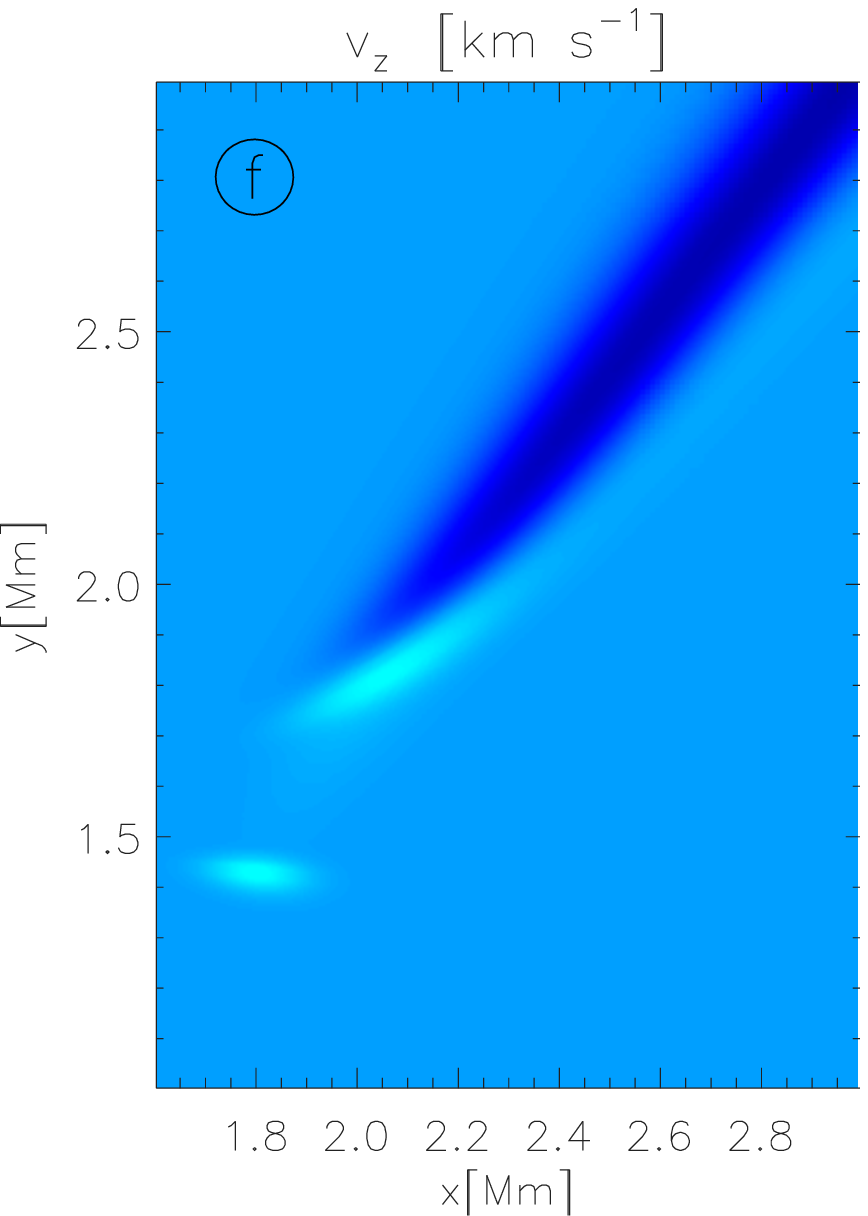}\\\vspace{0.2cm}\hspace{0.9cm}
           \includegraphics[width=11.6cm,angle=0]{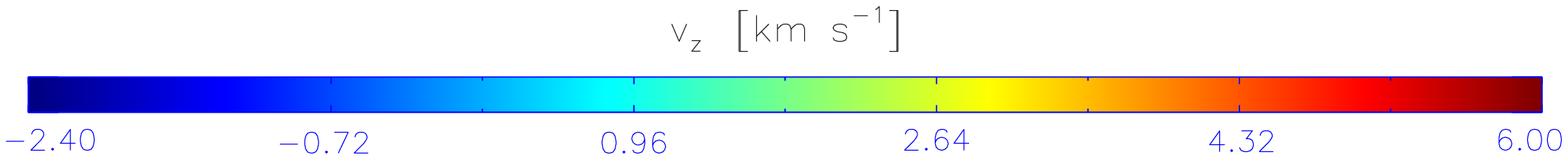}
          }
\caption{\small 
         Spatial profiles of the velocity $V_{\rm z}(x,y,t)$
         near the point of initial perturbation
         for the case of $x_{\rm 0}=2$~Mm, $w_{\rm x}=0.1$~Mm
         at $t=5$ s, $t=8$ s, $t=11$ s, $t=14$ s, $t=18$ s and $t=21$ s, 
	 presenting propagation through the transition region and
         evolution of the initial pulse.
	 A full-colour version of above figure and movie is available at www.pchmiel.republika.pl/store/Fig4.avi.
	}
\label{fig:pass}
\end{figure*}
%

%
\begin{figure}
\centering{
           \includegraphics[width=8.6cm,angle=0]{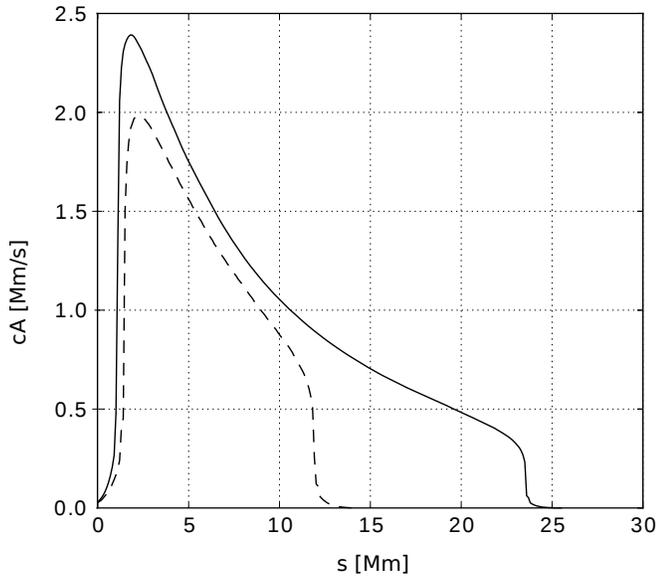}
          }
\caption{\small
         The Alfv\'en speed, $c_{\rm A}$, along the magnetic line,
  	 which crosses the point $(x_{\rm 0}, y_{\rm 0}) = (2~$Mm$, 1.75~$Mm$)$ (solid line) and
	 the point $(x_{\rm 0}, y_{\rm 0}) = (3~$Mm$, 1.75~$Mm$)$ (dashed line).
        }
\label{fig:cas}
\end{figure}
%

%
\begin{figure*}
\centering{
           \includegraphics[width=12.2cm,angle=0]{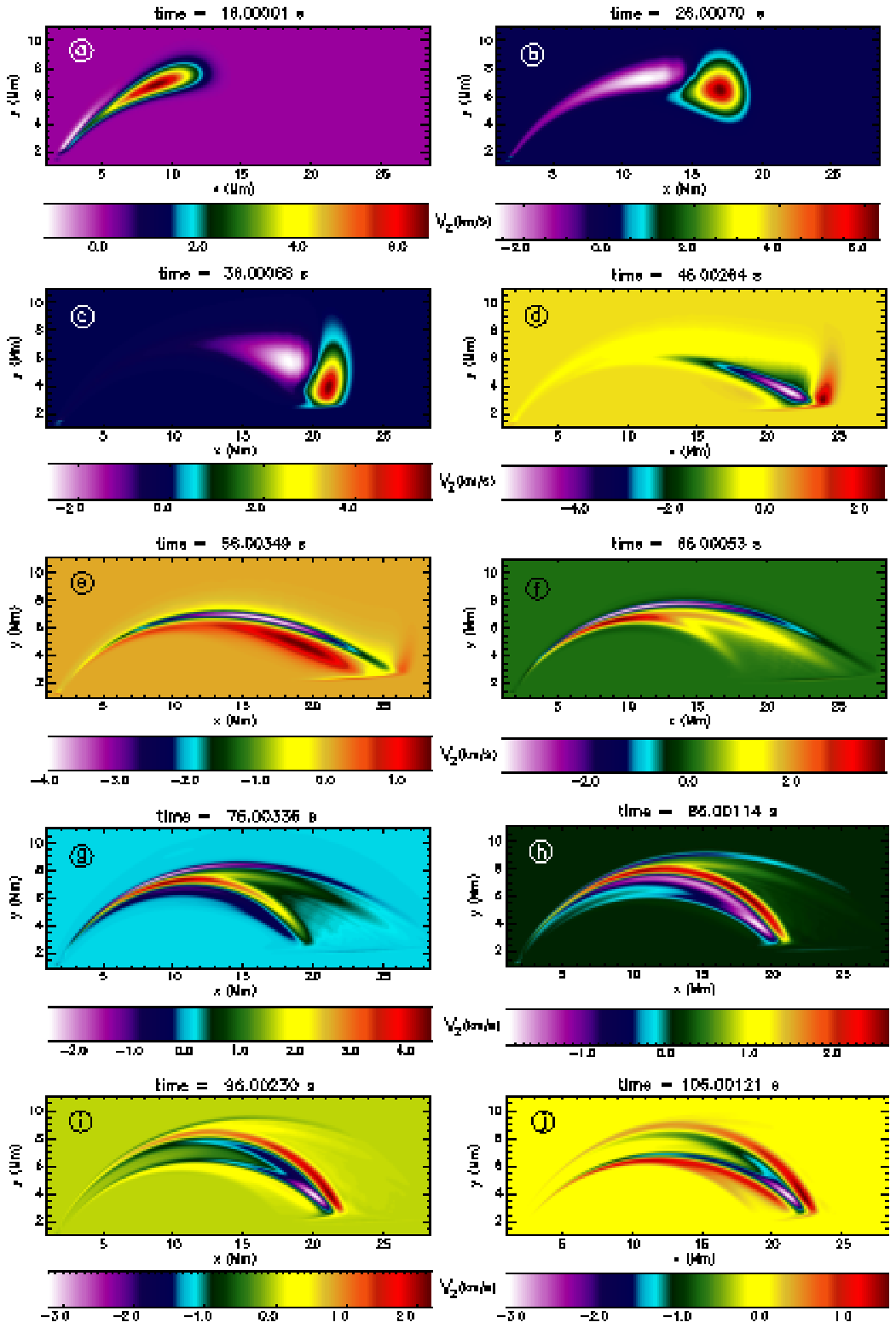}
          }
\caption{\small Spatial profiles of transverse velocity $V_{\rm z}(x,y,t)$
         for $x_{\rm 0}=2$~Mm, $w_{\rm x}=0.2$~Mm
         at $t=16$ s , $t=26$ s, $t=36$ s, $t=46$ s, $t=56$ s, $t=66$ s, $t=76$ s, $t=86$ s, $t=96$ s and $t=106$ s 
         (from top-left to bottom-right).
         A full-colour version of above figure and movie is available at www.pchmiel.republika.pl/store/Fig6.avi.
        }
\label{fig:6panels_2Mm}
\end{figure*}

%

%
In this model the Alfv\'en speed, $c_{\rm A}$, varies in both the $x$ and $y$ 
directions and it is expressed as follows: 
%
\beq
\label{eq:ca}
c_{\rm A}(x,y) = \frac{ | {\bf B}_{\rm e}(x,y) | }{\sqrt{\mu \varrho_{\rm e}(y)}}\, .
\eeq
%
Its profile is displayed 
in Fig.~\ref{fig:TeAlf} (bottom panel).
Note that the Alfv\'en speed 
is non-isotropic; higher values are located near $x_{\rm 0}=0$~Mm,
where the magnetic field is stronger,
while $c_{\rm A}$ decreases with larger values of $x_{\rm 0}$.
In the chromosphere $c_{\rm A}(x=0\, {\rm Mm},y=1.75\ 
{\rm Mm})$ is about $25$ km s$^{-1}$.  The Alfv\'en speed rises abruptly through 
the solar transition region, reaching a value of $c_{\rm A}(x=0\, {\rm Mm},y=10\, 
{\rm Mm}) = 10^{3}$ km s$^{-1}$ (Fig.~\ref{fig:TeAlf}, bottom panel). 
The increase of $c_{\rm A}(x=0~{\rm Mm},y)$ with height results from a faster decrease of $\varrho_{\rm e}(y)$
than  $B_{\rm e}(x=0\, {\rm Mm},y)$ with the atmospheric height.
%
%
\section{Numerical simulations of MHD equations}\label{sec:num_sim_MHD}
%
%
To solve Eqs (\ref{eq:MHD_rho})-(\ref{eq:MHD_CLAP}) numerically, we use 
the FLASH code (Fryxell et al. 2000; 
 Lee \& Deane 2009; Lee 2013),
in which a third-order unsplit Godunov-type solver with various slope limiters 
and Riemann solvers as well as Adaptive Mesh Refinement (AMR) (MacNeice et 
al. 1999) are implemented.  The minmod slope limiter and the Roe Riemann 
solver (e.g., T\'oth 2000) are used. We set the simulation box as
$(1\, {\rm Mm},28.5\, {\rm Mm}) \times (1\, {\rm Mm},11\, {\rm Mm})$ 
and impose fixed in time boundary conditions for all plasma quantities
in the $x$- and $y$-directions, while all plasma quantities remain invariant 
along the $z$-direction; however, note that both $V_{\rm z}$ and $B_{\rm z}$ 
differ from zero.

In our present work, we use a static, non-uniform grid with a minimum (maximum) level 
of refinement set to $4$ ($6$). 
We performed 
the grid convergence studies by refining this by a factor of two. 
As the numerical results remained essentially similar for the grid of maximum blocks levels 6 and 7, we adopted the 
former to get the results presented in this paper. 
Note that small size blocks of numerical grid occupy the solar transition region
and the region located along Alfv\'en wave propagation path 
(Fig.~\ref{fig:blk}), and every numerical block consists of $8\times 8$ identical 
numerical cells.  This results in an excellent resolution of steep spatial 
profiles and greatly reduces the numerical diffusion in these regions. 
%
\subsection{Initial perturbations}
%
%
We perturb initially (at $t=0$ s) the equilibrium described in Sec.~\ref{sec:equil} 
by a Gaussian pulse in the $z$-component of velocity given by 
%

%
\beq
\label{eq:init_per}
V_{\rm z}(x,y,t=0) = A_{\rm v} e^{ -\frac{(x-x_{\rm 0})^2}{w_{\rm x}^2}
					  -\frac{(y-y_{\rm 0})^2}{w_{\rm y}^2} } \, ,
\eeq
%
where $A_{\rm v}$ is the amplitude of the pulse, $(x_{\rm 0},y_{\rm 0})$ is its 
initial position and $w_{\rm x}$ and $w_{\rm y}$ denote its widths
along $x$- and $y$-directions, respectively.  We set and 
hold fixed $y_{\rm 0}=1.75$~Mm, $A_{\rm v}=3$ km~s$^{-1}$ and $w_{\rm y}=0.1$~Mm,
but allow $x_{\rm 0}$ and $w_{\rm x}$ to vary.
This shows that Alfv\'en waves in our model are generated in the solar chromosphere
just above the photosphere, 
which makes our model significantly different than that considered by Del Zanna 
et al. (2005) {and Miyagoshi et al.~(2004)}, who launched the initial pulses in the solar corona.  
Physical consequences of these differences are described in Sect.~4.  
It is clear that 
our more realistic model significantly extends the model considered by Del 
Zanna et al. (2005) and provides a platform to pursue the modelling of Alfv\'en 
wave phase-mixing in realistic solar atmosphere where an appropriate driver 
(dynamical phenomena in the solar chromosphere) is responsible for excitation 
of such waves.

%
Note that in our 2.5D model, the Alfv\'en waves decouple from magnetoacoustic 
waves and it can be described solely by $V_{\rm z}(x,y,t)$.  As a result, the 
initial pulse triggers Alfv\'en waves that in the linear limit are approximately 
described by the following wave equation:
%
\beq
\label{eq:Vz_linear}
\frac{\partial^2 V_{\rm z}}{\partial t^2} = c_{\rm A}^2(x,y) \frac{\partial^2 
V_{\rm z}}{\partial s^2}\, ,
\eeq
%
where $s$ is the coordinate along a magnetic field line.

%
\section{Results of numerical simulations}\label{sec:resultsOFnumSIM}
%

We simulate small amplitude and impulsively excited Alfv\'en waves and investigate
their propagation along magnetic field lines 
which are parallel to magnetic vectors 
shown in Fig.~\ref{fig:mag}.
This magnetic field configuration mimics an 
asymmetric solar arcade,
 which is more realistic than that considered by Del Zanna et al. (2005). 

%
\begin{figure*}
\centering{
           \includegraphics[width=6.1cm,angle=0]{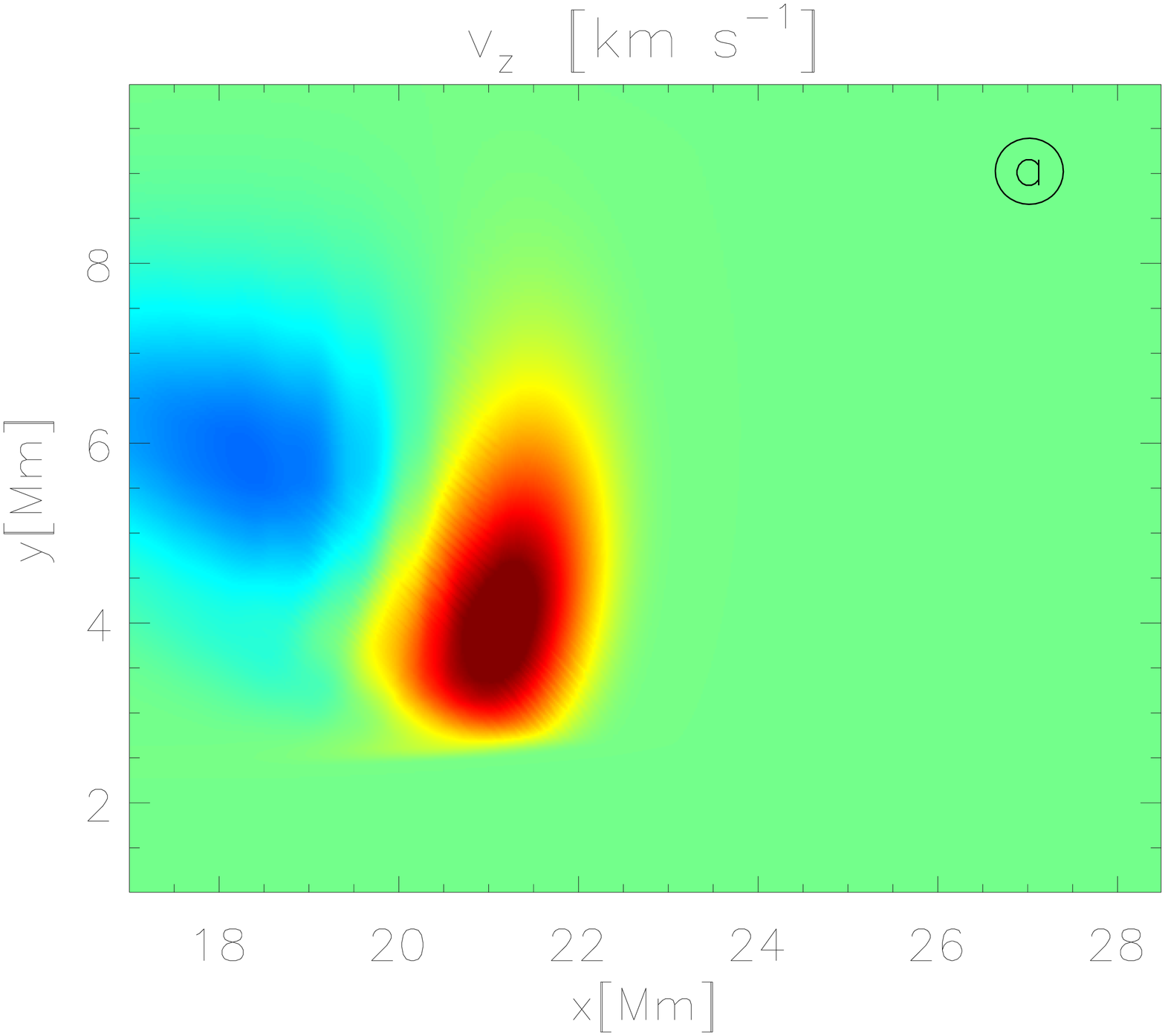}
           \includegraphics[width=6.1cm,angle=0]{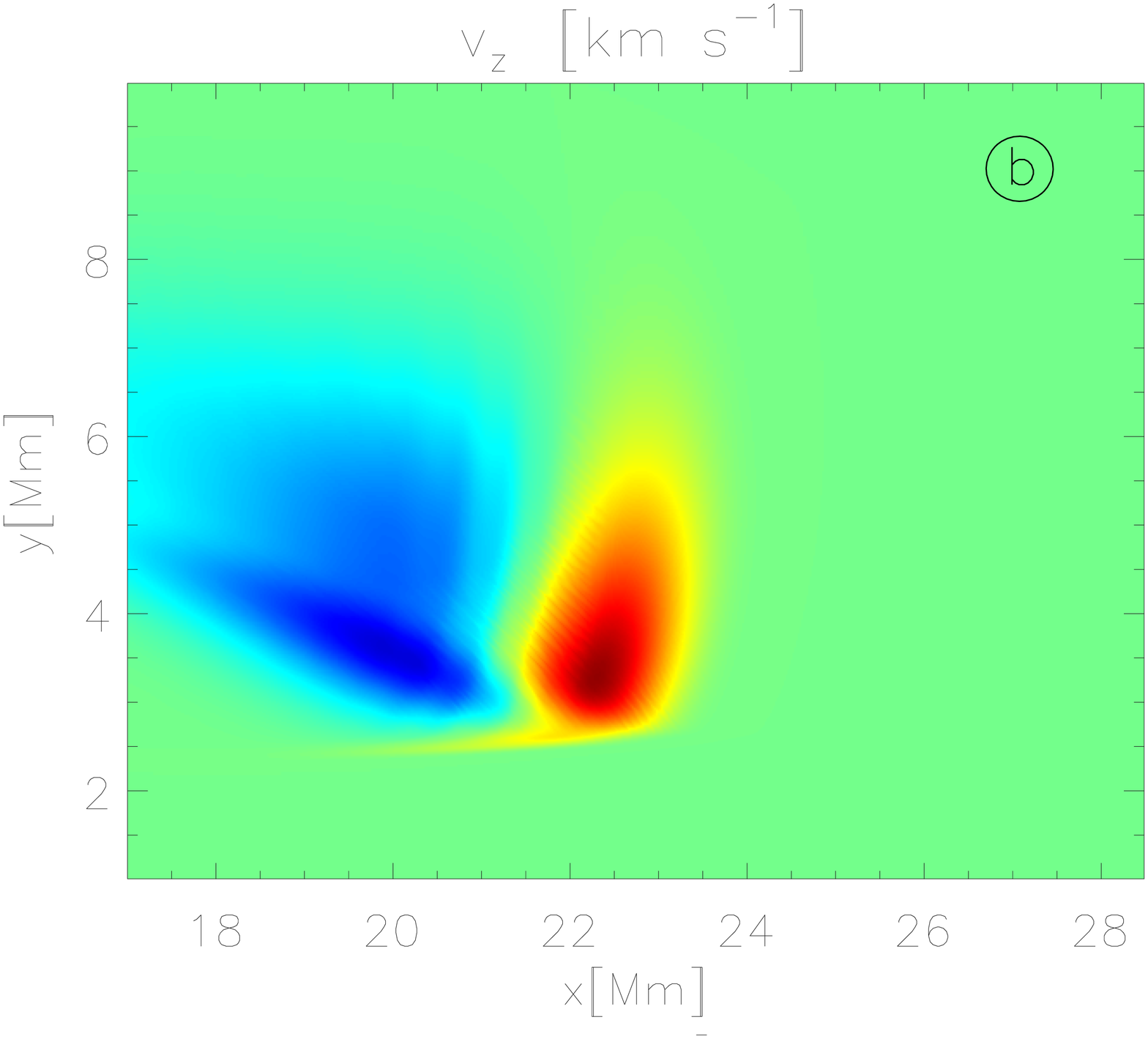}\\
           \includegraphics[width=6.1cm,angle=0]{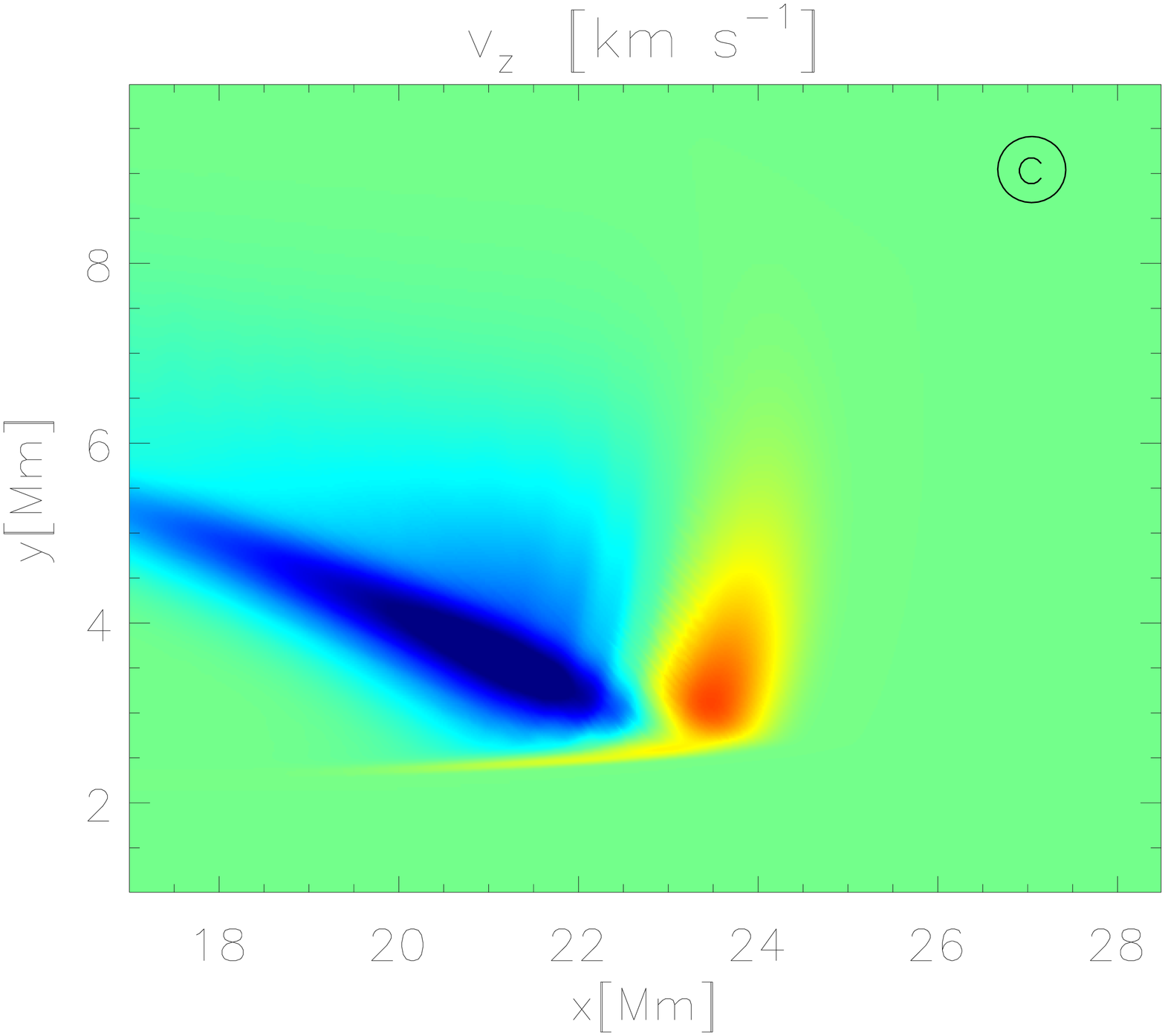}
           \includegraphics[width=6.1cm,angle=0]{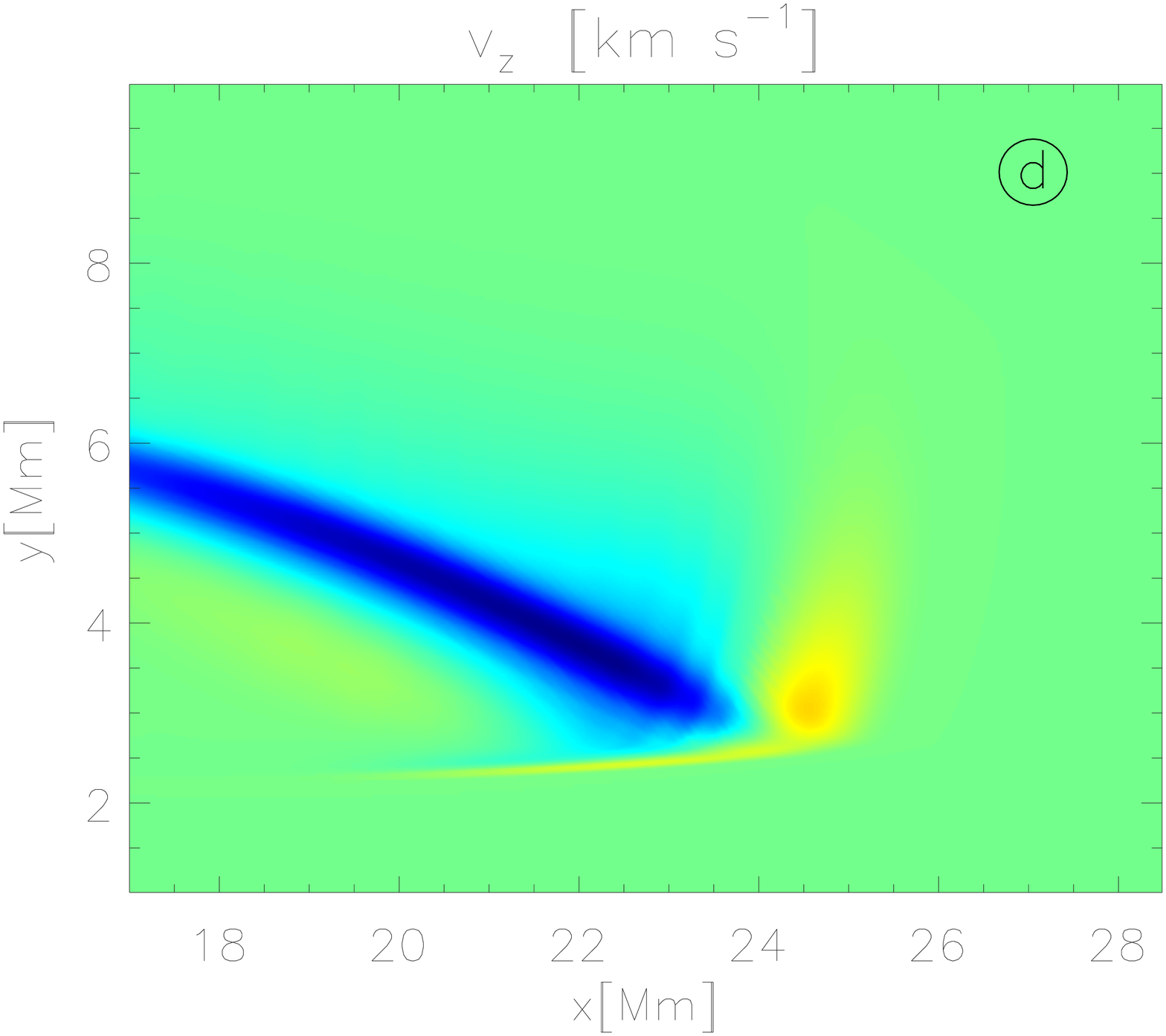}\\
           \includegraphics[width=6.1cm,angle=0]{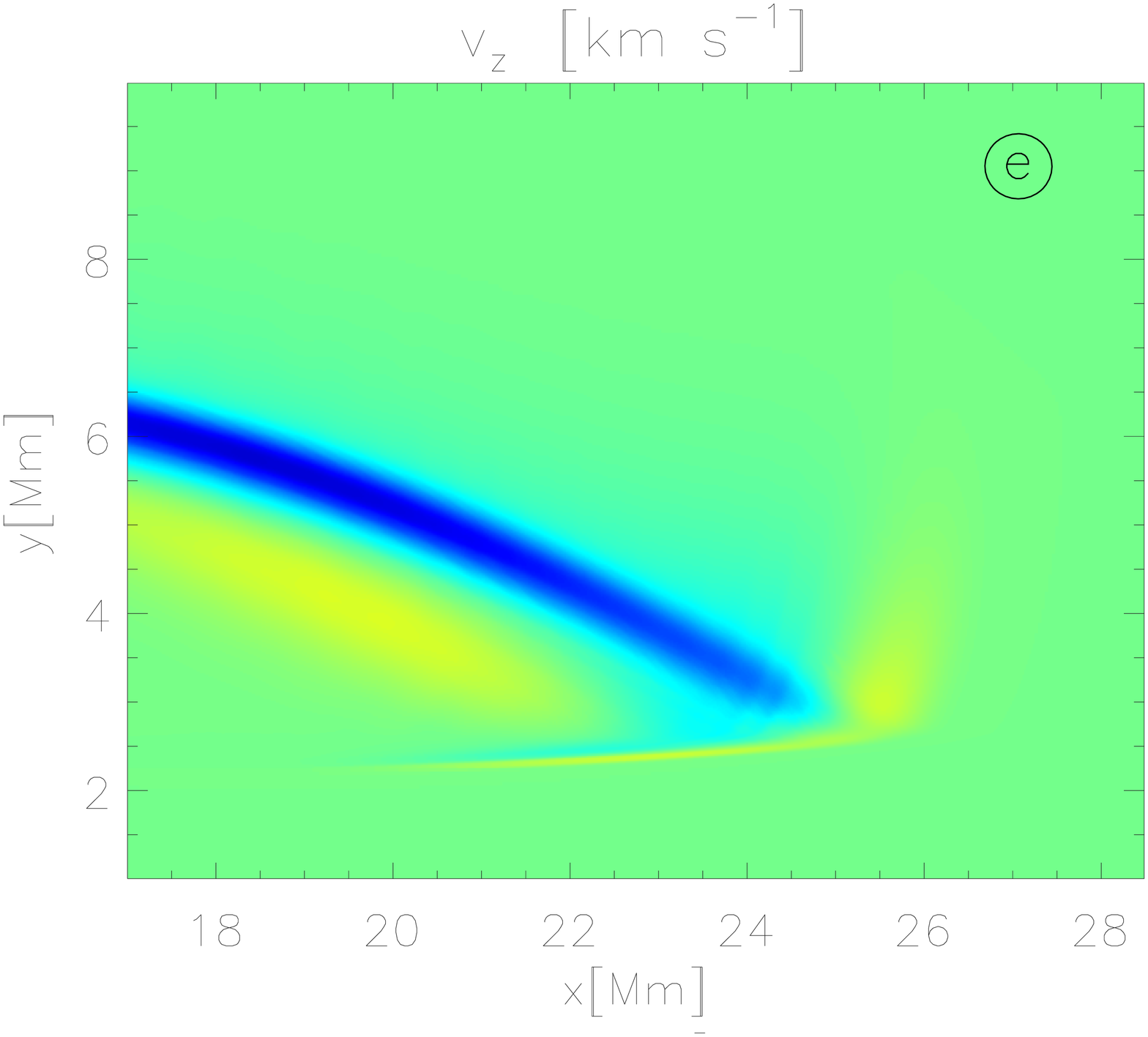}
           \includegraphics[width=6.1cm,angle=0]{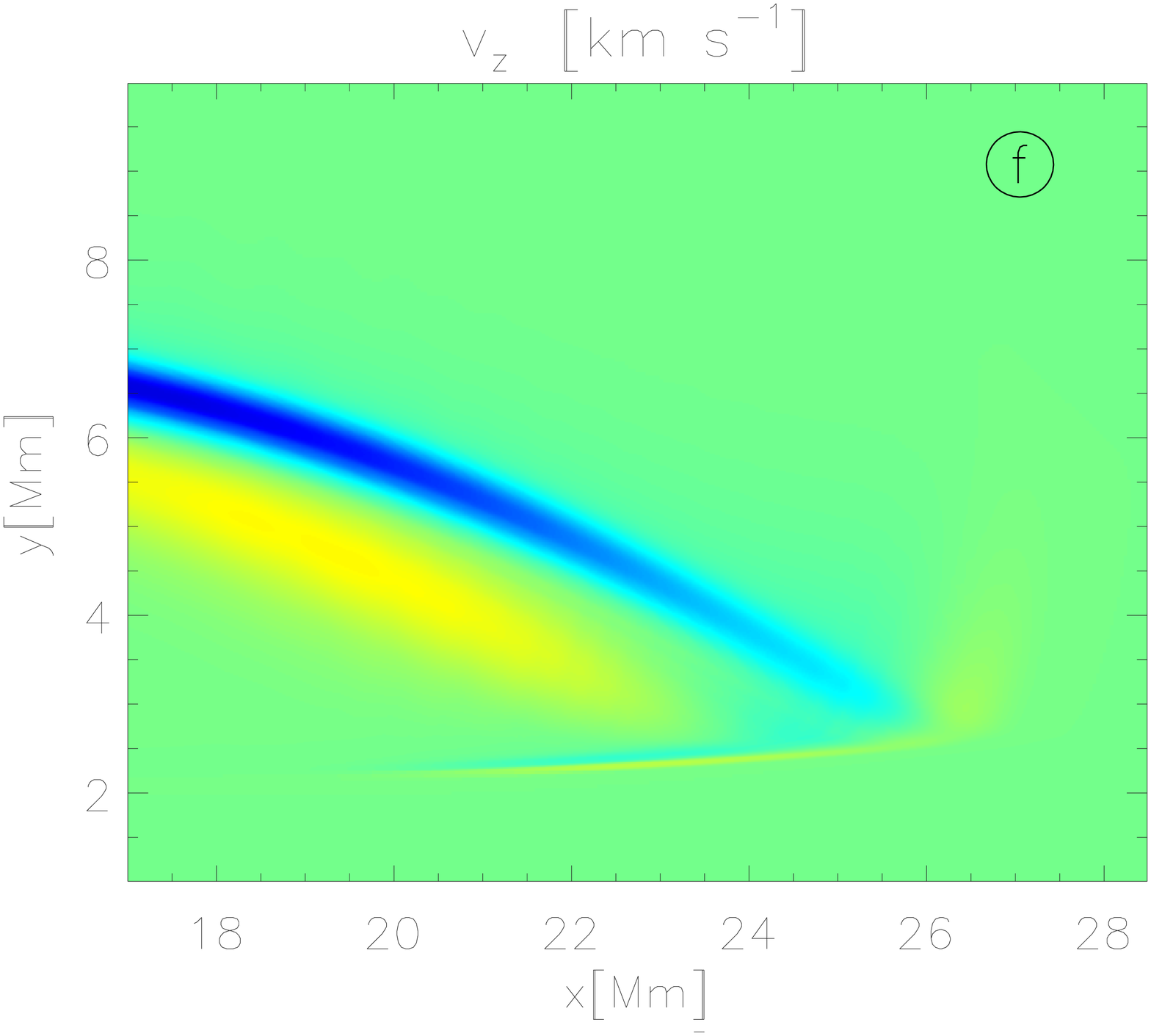}\\
           \includegraphics[width=6.1cm,angle=0]{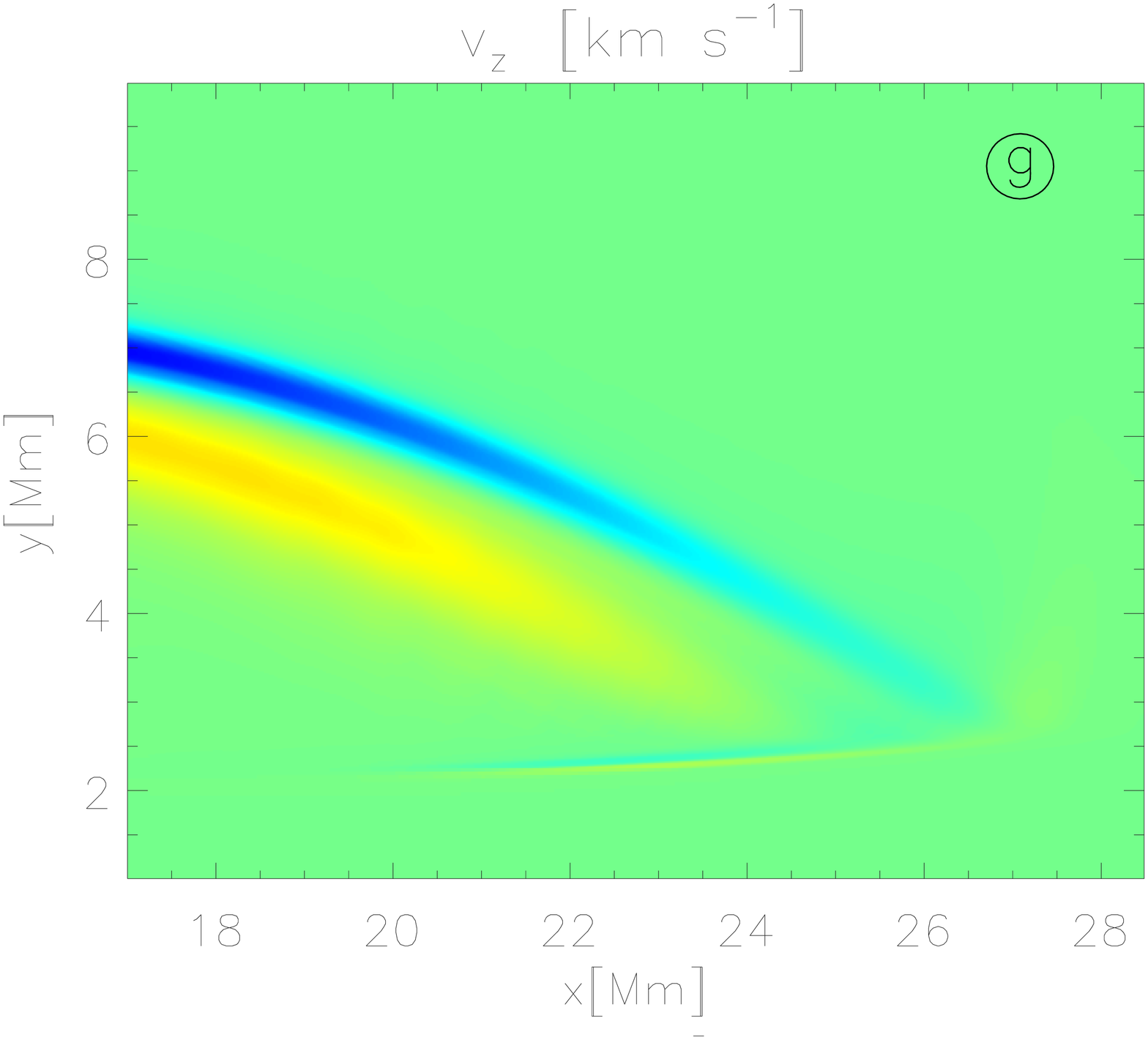}
           \includegraphics[width=6.1cm,angle=0]{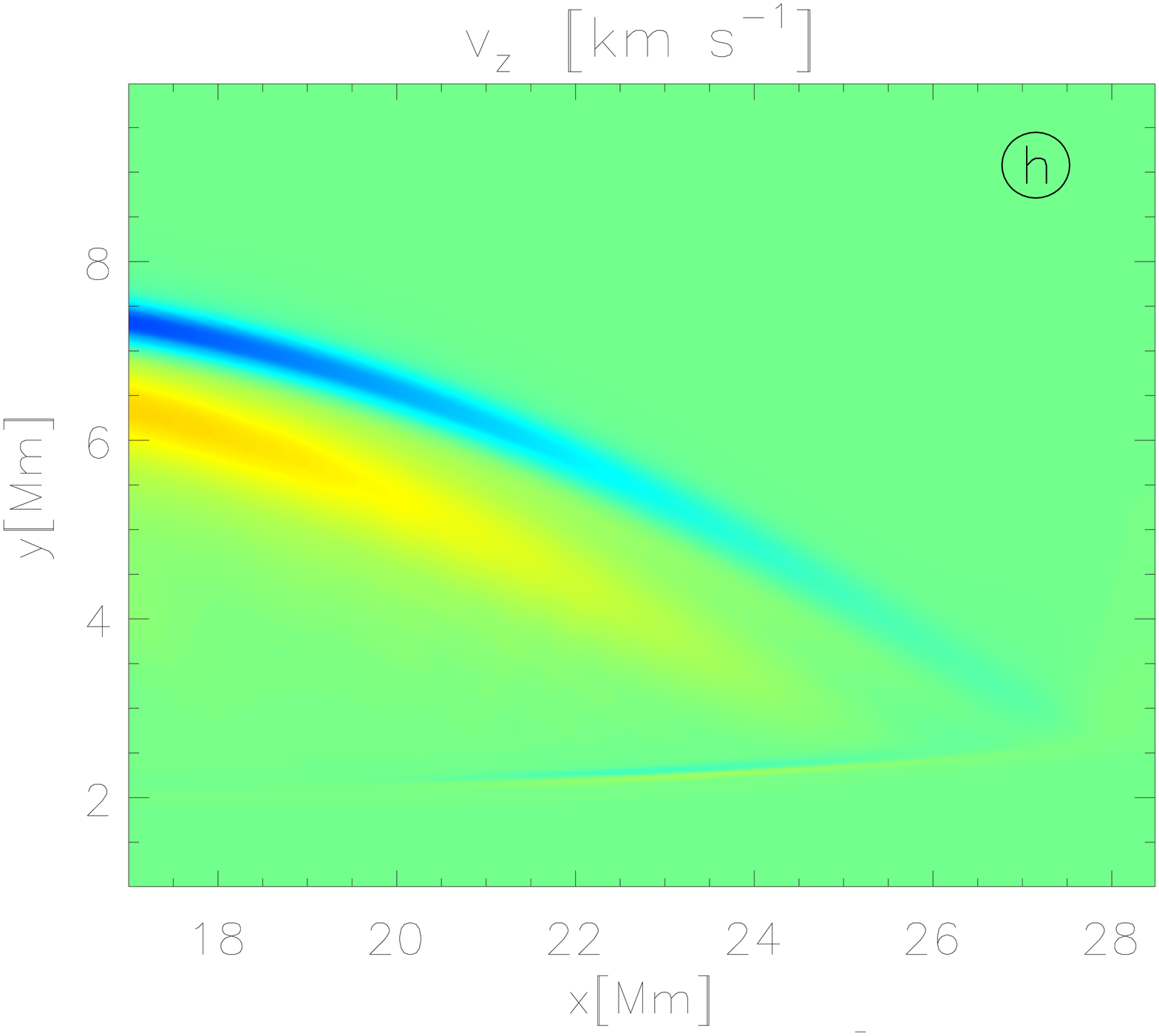}\\\hspace{0.9cm}
           \includegraphics[width=12.2cm,angle=0]{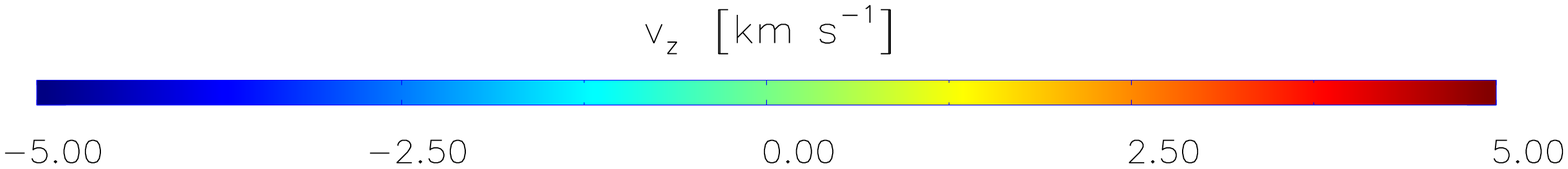}
          }
\caption{\small
         Spatial profiles of transverse velocity $V_{\rm z}(x,y,t)$
         for $x_{\rm 0}=2$~Mm, $w_{\rm x}=0.2$~Mm
         at $t=36$ s, $t=40$ s, $t=44$ s, $t=48$ s, $t=52$ s, $t=56$ s, $t=60$ s and $t=64$ s 
         (from top-left to bottom-right),
	 illustrating a partial reflection of Alfv\'en waves from the transition region.
        }
\label{fig:reflection}
\end{figure*}
%

%
\begin{figure*}
\centering{
           \includegraphics[width=12.2cm,angle=0]{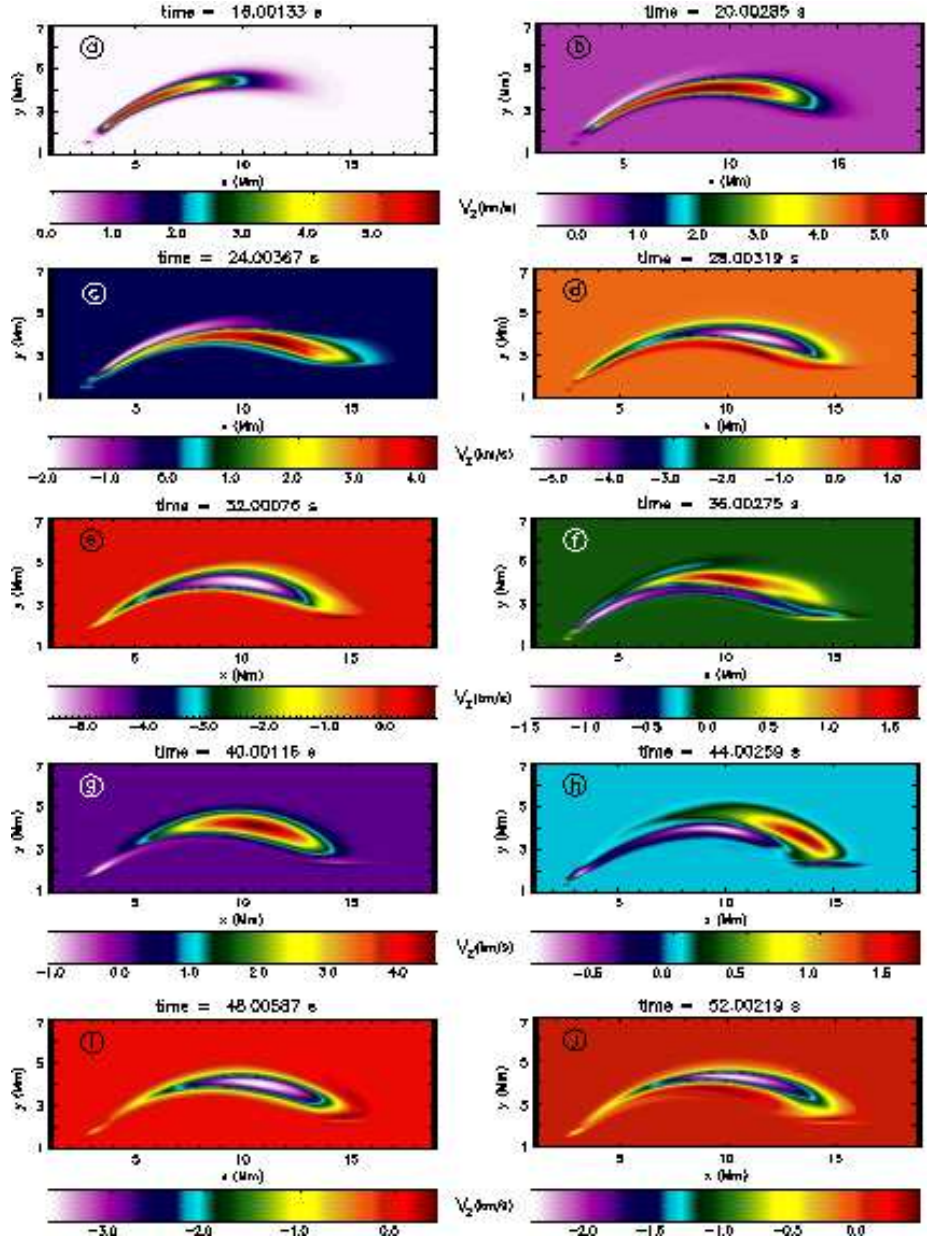}
          }
\caption{\small
         Spatial profiles of transverse velocity $V_{\rm z}(x,y,t)$
         for $x_{\rm 0}=3$~Mm, $w_{\rm x}=0.2$~Mm
         at $t=16$ s, $t=20$ s, $t=24$ s, $t=28$ s, $t=32$ s, $t=36$ s, $t=40$ s, $t=44$ s, $t=48$ s and $t=52$ s 
         (from top-left to bottom-right). 
         A full-colour version of above figure and movie is available at www.pchmiel.republika.pl/store/Fig8.avi.
        }
\label{fig:6panels_3Mm}
\end{figure*}
%

%
\begin{figure}
\centering{
           \includegraphics[width=6.6cm,angle=0]{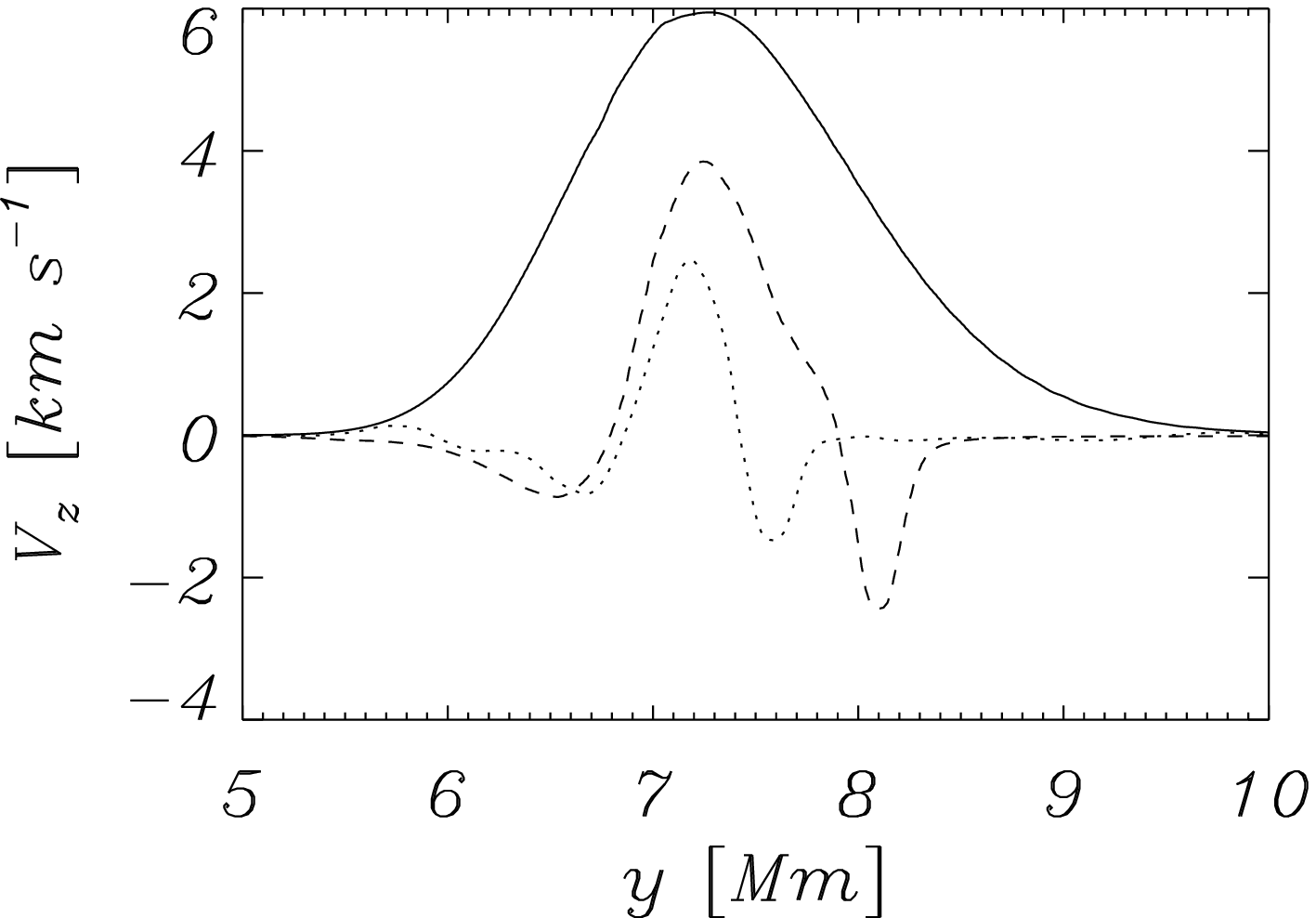}\\
           \includegraphics[width=6.6cm,angle=0]{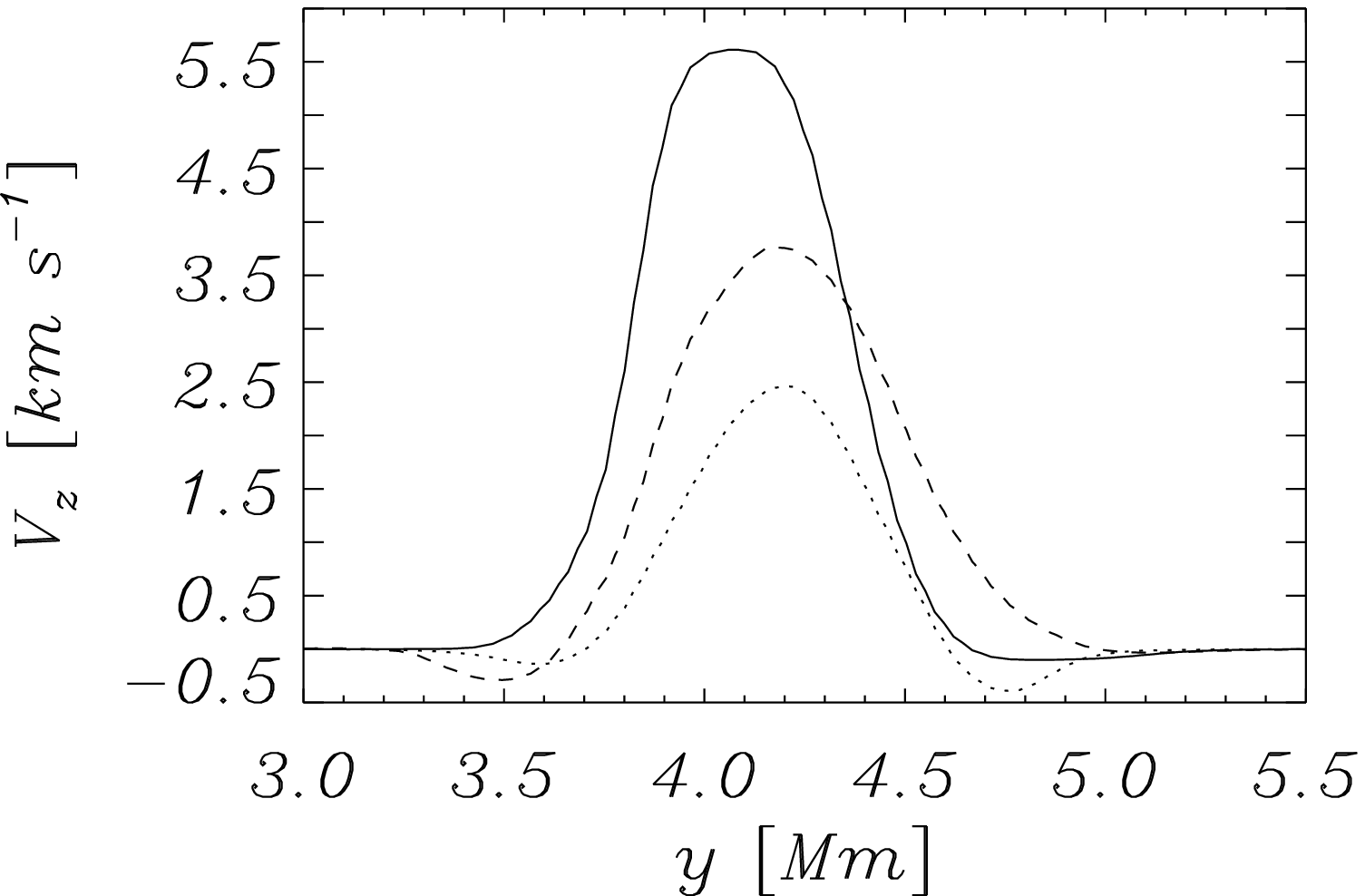}
          }
\caption{\small
         Vertical profiles of $V_{\rm z}(x=12\, {\rm Mm})$ for the case of $x_{\rm 0}=2\, {\rm Mm}$ (top panel)
	 at $t=22$~s (solid line), $t=74$~s (dashed line) and $t=124$~s (dotted line),
	 and $V_{\rm z}(x=9\, {\rm Mm})$ for the case of $x_{\rm 0}=3\, {\rm Mm}$ (bottom panel)
	 at $t=20$~s (solid line), $t=40$~s (dashed line) and $t=60$~s (dotted line). 
        }
\label{fig:cross-section}
\end{figure}
%
%
\begin{figure}
\centering{
           \includegraphics[width=6.2cm,angle=0]{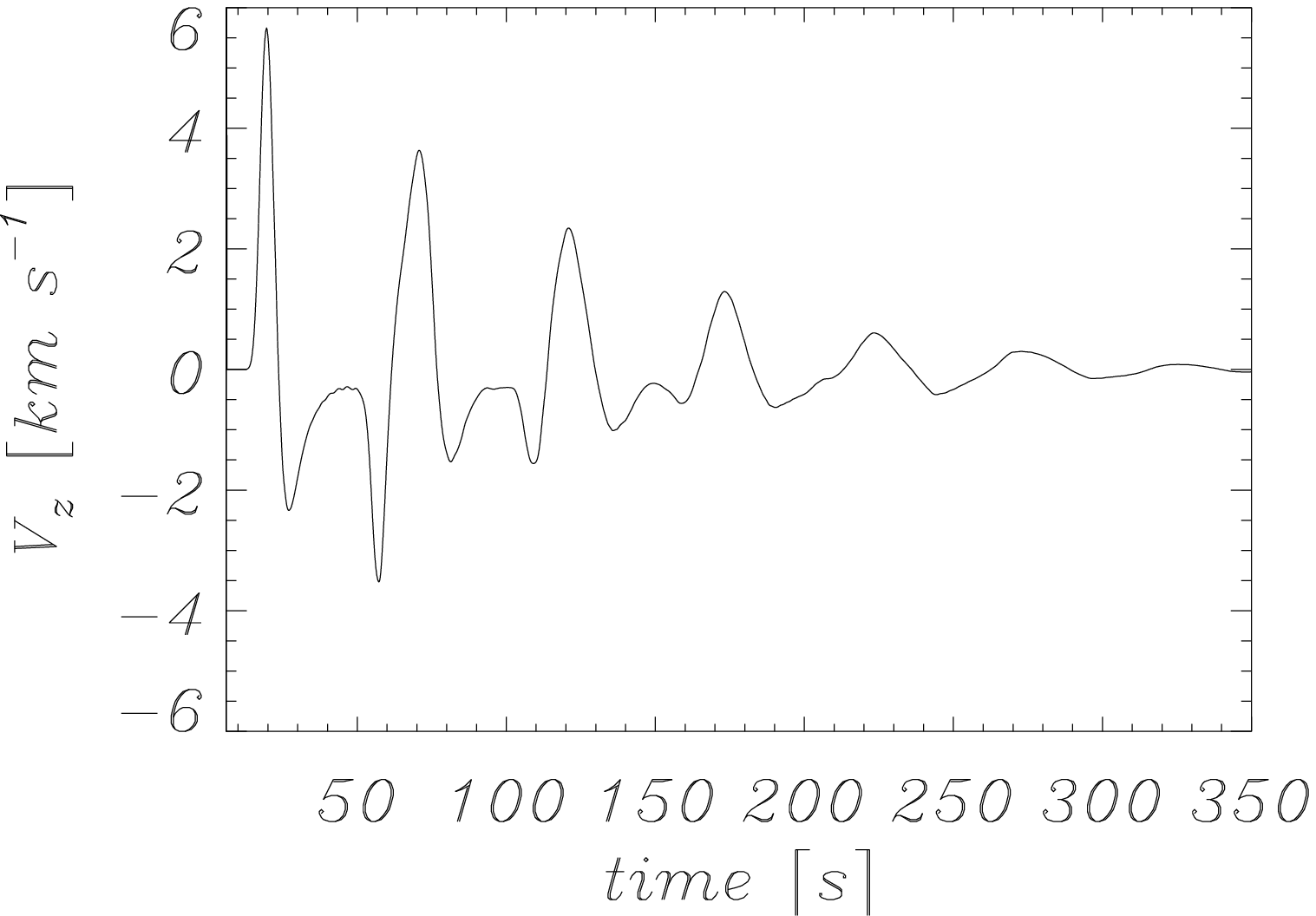}\\
           \includegraphics[width=6.2cm,angle=0]{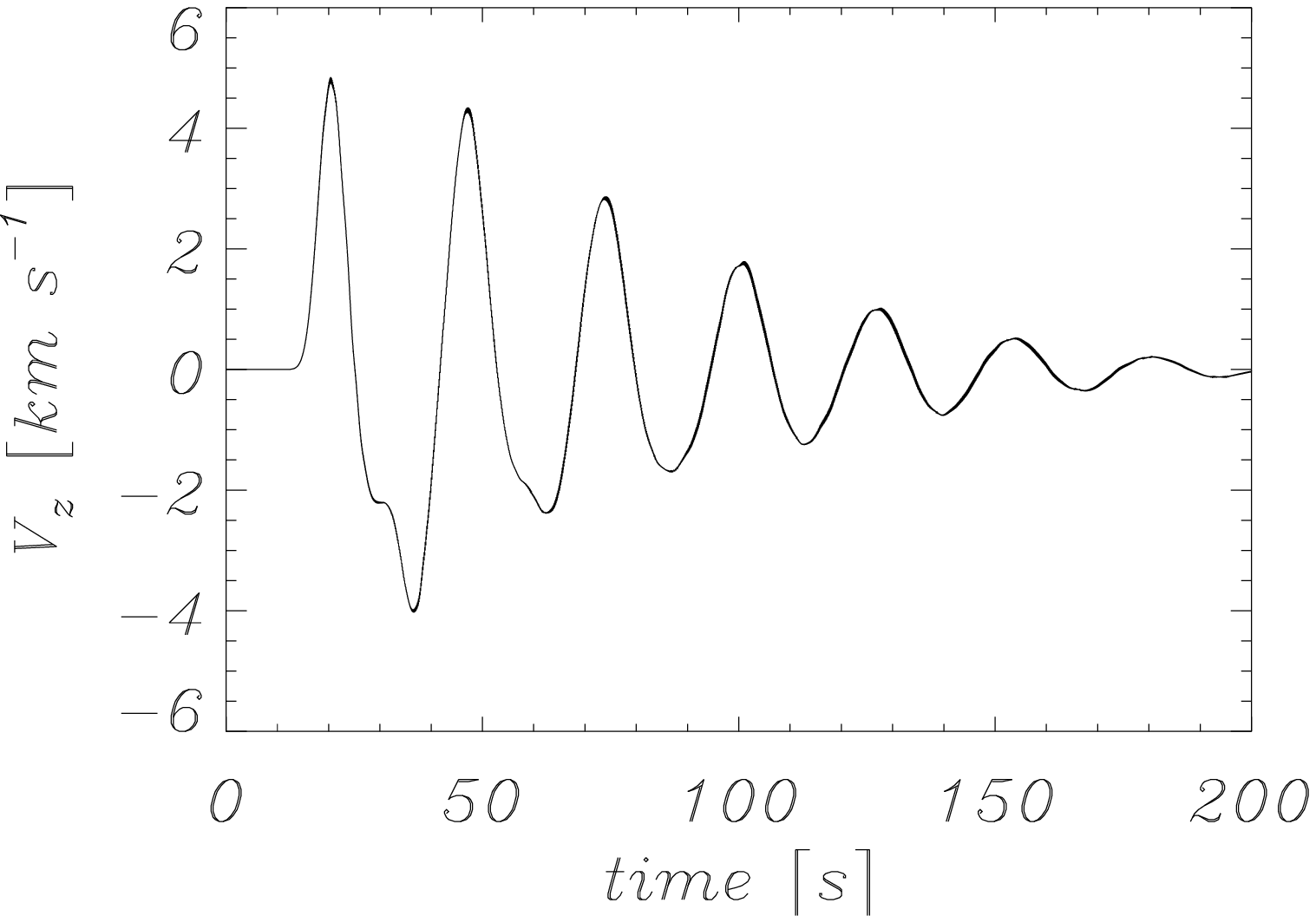}\\
           \includegraphics[width=6.2cm,angle=0]{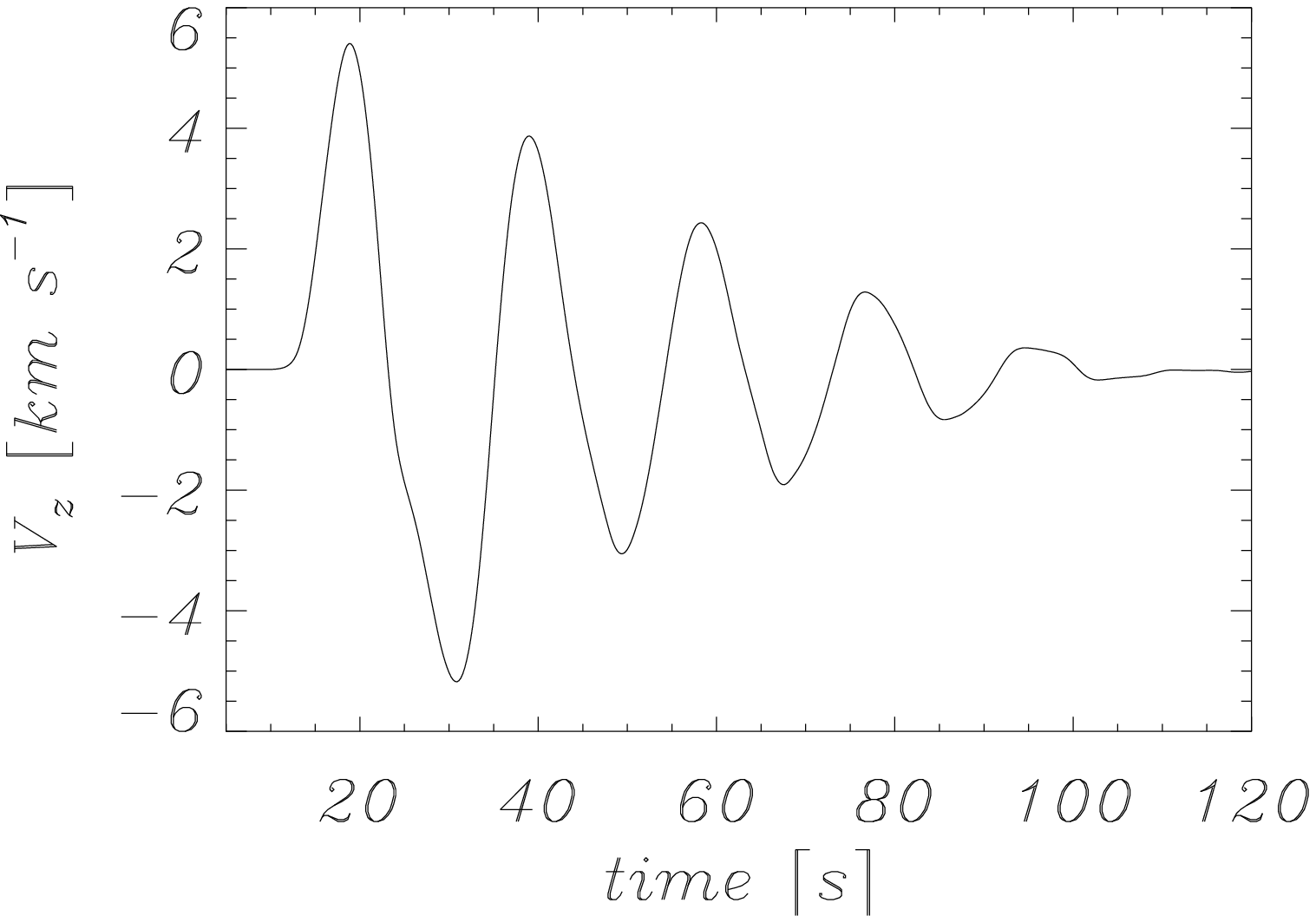}
          }
\caption{\small
         Time-signatures of $V_{\rm z}$
         for $w_{\rm x}=0.2$~Mm, $x_{\rm 0}=2$ Mm (top panel),
	 $x_{\rm 0}=2.5$ Mm (middle panel) 
         and $x_{\rm 0}=3$ Mm (bottom panel).
         Detection points are located at ($x=12$ Mm, $y=7$ Mm) 
         (top panel), ($x=11$ Mm, $y=5.1$ Mm) (middle panel)
	 and ($x=9$ Mm, $y=4.2$ Mm) (bottom panel). 
        }
\label{fig:ts}
\end{figure}
%

%
\subsection{Initial stage of wave propagation}
%
 The Gaussian pulse of initial perturbation in the $z$-component of velocity,
 described by Eq.~(\ref{eq:init_per}),
 decouples into two moving in opposite directions pulses.
 The downwardly propagating pulse fades very fast,
 however the second one propagates through the chromosphere and transition region to the solar corona (Fig.~\ref{fig:pass}).

As a result of several effects such as finite size of the Alfv\'en wave
pulse, inclination of the magnetic field, and highly inhomogeneous Alfv\'en 
speed, $c_{\rm A}(x,y)$, the wave signal in our numerical simulations is 
deformed and elongated while passing in the arcade through the transition 
region.  In Fig.~\ref{fig:pass}, we observe the Alfv\'en waves reaching the 
transition region located at $y\simeq 2.7$~Mm, and accelerated due to a 
sudden increase of local Alfv\'en speed (panels $a$ and~$b$).  The wave 
signal penetrates gradually through the transition region. {First, 
the initially circular wave profile is distorted to ellipsoidal shape
because Alfv\'en speed slightly raise along the inclined magnetic field 
lines (Fig.~\ref{fig:cas}), and next becomes significantly elongated 
above the transition region (Fig.~\ref{fig:pass}, panel $c$) due to 
sudden increase of the Alfv\'en speed (Fig.~\ref{fig:cas}).
}

This clearly shows that the upper part of the wave signal,
propagating along the magnetic field lines,
penetrates the solar corona earlier
than the middle and bottom parts of the signal (Fig.~\ref{fig:pass}, panel $b$).
Moreover, the upper part of the wave follows along different (upper)
magnetic field line than the right part of the wave that propagates over the lower
magnetic field line, which results in deformation of the initial pulse and finally intensifies 
the phase-mixing that can be seen at the panels $d$ and~$e$ of Fig.~\ref{fig:pass}.
Here, the upper part of the wave signal
located at the upper magnetic field lines
illustrated as a white patch (Fig.~\ref{fig:pass}, panels $d$ and~$e$)
is in a different phase than the signal located at the lower
magnetic field line, that already passed the transition region.
This additional process of the wave deformation at the transition region
significantly influences Alfv\'en wave propagation in the curved coronal arcades,
introducing different phases of the wave signal in the arcade at beginning stage.
Notice, that the Alfv\'en waves suffer from the partial reflection from the transition region
and a small signal propagates backward into the solar surface,
which can be spotted at $(x,y)\simeq(2$~Mm$,1.8$~Mm$)$ in Fig.~\ref{fig:pass}, panels $e$ and~$f$.
%
%
\subsection{Different spatial positions of initial pulses}
%
%
In our approach for a fixed value of $y_0$, the length and curvature of magnetic 
field lines vary with 
$x_{\rm 0}$. For a smaller value of $x_{\rm 0}$ a magnetic line is longer, 
more curved, and less inclined to the horizontal direction
(Fig.~\ref{fig:mag}) as well as Alfv\'en speed $c_{\rm A}(x,y)$ experiences a larger 
structuring there (Fig.~\ref{fig:TeAlf}, bottom panel).
While a larger value of $x_{\rm 0}$ corresponds to a shorter, 
less curved and more inclined magnetic field lines as well as less varying $c_{\rm A}(x,y)$. 
Here, we consider two different cases of initial 
pulse position: $x_{\rm 0}=2$~Mm and $x_{\rm 0}=3$~Mm.  
In the first case, 
the Alfv\'en waves propagate higher up along longer and more curved
magnetic field lines. 
However, in the case of $x_{\rm 0}=3$~Mm, the inclination of 
magnetic field lines to $x$-axis is much larger and this results in 
approximately two times smaller size of the arcade with less curved field lines.
In Fig.~\ref{fig:cas}, we notice that $c_{\rm A}(s)$ varies less along the magnetic field line, $s$,
for a small arcade, reaching a value of $2$~Mm~s${}^{-1}$ (dashed line),
while for a large arcade it attains $2.4$~Mm~s${}^{-1}$ (solid line).
Here, a value of $s=0$ corresponds to the point at which the initial perturbation 
was launched, $(x_{\rm 0}, y_{\rm 0})$, 
equal to $(2$~Mm$,1.75$~Mm$)$ for the large arcade (solid line)
and $(3$~Mm$,1.75$~Mm$)$ for the small arcade (dashed line).
%
%
\subsubsection{Large arcade}
%
%
Our numerical results are presented in Fig.~\ref{fig:6panels_2Mm}, which illustrates 
spatial profiles of Alfv\'en waves generated by the initial Gaussian pulse 
given by Eq.~(\ref{eq:init_per}). 
This pulse triggers counter-propagating Alfv\'en waves
(e.g., Murawski \& Musielak 2010). The wave that propagates downward 
enters a region of small Alfv\'en speed, which results from high mass density 
there. The presented results show that the evolution of
the downwardly propagating waves (Fig.~\ref{fig:pass}) is much 
less dynamic than the evolution of Alfv\'en waves that already reach the upper 
regions of the solar atmosphere.
The upwardly propagating Alfv\'en waves, reaching the transition region, accelerate due to increase of local Alfv\'en speed
(Fig.~\ref{fig:TeAlf}, bottom panel, and Fig.~\ref{fig:cas}, solid line)
and penetrate into the solar corona.
In Fig.~{\ref{fig:6panels_2Mm}}, panel $a$, the wave signal arrived to the 
level $y=9$~Mm.
At $t=26$ s (panel $b$), the Alfv\'en waves are at the apex
while fanning its spatial profile as a result of diverging magnetic field lines (Fig.~\ref{fig:mag}),
and the spatial variation of the Alfv\'en speed (Fig.~\ref{fig:TeAlf}, bottom panel).
This process becomes even more pronounced in time as the Alfv\'en waves penetrate 
the regions of more diverged field lines. 
Behind the Alfv\'en wave signal, there is visible fanning of 
a negative value of $V_{\rm z}$ (violet-white patch),
that is a remnant after the Alfv\'en wave passing through the plasma medium 
(Murawski \&~Musielak 2010).
Now, the violet patch seen in Fig.~\ref{fig:6panels_2Mm} becomes elongated and it contains a white 
subregion located at $x=12$~Mm, $y=7$~Mm (panel $b$).
This white 
subregion corresponds to negative ($V_{\rm z}\approx-2.1$ km s${^{-1}}$)
values of transversal velocity while nearby laying positive velocity ($V_{\rm z}\approx6.4$ km s${^{-1}}$) 
region is represented by a red patch.
This results from the finite-size of the initial 
pulse whose left and right-sides have to pass significantly different distances, 
because the magnetic field line along the left-side path is longer than the magnetic 
field line along the right-side in the considered arcade.
%
Moreover, a strong fanning of the wave signal spatial profile
is a consequence of
the Alfv\'en speed, $c_{\rm A}(x,y)$, which attains different values along 
different 
magnetic interfaces (Fig.~\ref{fig:TeAlf}, bottom panel). 
At $t=36$ s, the already downwardly 
propagating Alfv\'en waves reach the transition region (panel $c$) and they undergo 
partial reflection from this region. 

The resulting wave reflection is well seen at the 
consecutive moments of time, e.g., at $t=46$~s (panel $d$) and at $t=56$~s (panel $e$), 
respectively.
However, at later moments of time the Alfv\'en waves profiles 
become more complicated;
we can notice a partial reflection of the wave signal with negative value of $V_{\rm z}$,
illustrated by narrow elongated violet-white patch (panels $e$ and~$f$).
At $t=66$~s the Alfv\'en waves reflect again at the left side, near $x_{\rm 0}=2$~Mm
and propagate with positive phase velocity (red patch) at $t=76$~s with amplitude $V_{\rm z}\approx4.2$ km s${^{-1}}$
in right direction.
We observe a long narrow signal seen as a red patch, which reaches the transition region
near $x_{\rm 0}=21$~Mm at $t=86$~s (panel $h$).
Reflections of the Alfv\'en waves are well seen at $t=86$~s (panel $h$) and~$t=96$~s (panel $i$) as violet-white patches 
near $x_{\rm 0}=21$~Mm.
Such reflections result in complex Alfv\'en waves spatial profiles.
We provide details on the partial reflection in Sect.~4.4,
where we evaluate the corresponding reflection coefficient.

Fig.~\ref{fig:reflection} presents in some details the Alfv\'en waves reflection.
The arriving waves (panel $a$, red patch) 
hit the transition region and start reflecting in there, decreasing its amplitude,
which is well seen in consecutive moments of time.
In panel $b$, the reflected, backwardly propagating wave is illustrated by a violet-white patch,
superimpose on the arriving negative phase waves.
The effect of superimposition of the wave signals
is significant, because of fanning of the wave spatial profiles,
which are a consequence of large arcade, where Alfv\'en speed gradient and diverged magnetic field lines are large.
At later moments of time (panels $e$ to $h$),
we also see the evolution of the lagging wave signals 
reflected from the transition region
as a red patch under the backward propagating main signal (violet-white patch).
Note that there is clearly seen small partially non-reflected (transmitted) Alfv\'en waves
propagating downwardly into the solar surface under the transition region.
These waves are in phase
and their amplitude is about $1$~km~s${}^{-1}$.
%

 It is worth to compare our results to those previously obtained by Del Zanna et al. 
 (2005), who considered an isothermal symmetric solar arcade with the transverse signal undergoing 
 reflection from the transition region. 
In our non-isothermal asymmetric arcade, the numerical results 
show how the Alfv\'en wave pulse is widened by the diverged magnetic field lines, becomes 
gradually reflected in the transition region, and creates a non-regular structure of 
the reflected wave signal (see Fig.~\ref{fig:6panels_2Mm}, bottom-right). Later on, 
the diverged magnetic field configuration, inhomogeneous Alfv\'en speed, difference of magnetic field lengths
and the reflected waves interaction with 
the ongoing wave train complicates even more the spatial velocity profiles in the 
arcade.

%

\subsubsection{Short arcade}

%
Spatial profiles of Alfv\'en waves,
which result from the initial Gaussian pulse given by Eq.~(\ref{eq:init_per}), 
with $x_{\rm 0}=3$~Mm and $w_{\rm x}=0.2$~Mm are presented in  
Fig.~\ref{fig:6panels_3Mm}.  Initial pulse triggers the counter-propagating 
Alfv\'en waves in a comparatively less curved magnetic arcade.
In this case, process of initial deformation of the wave pulse during propagation throughout the transition region
is more effective because magnetic field lines are more horizontal than in the case of $x_{\rm 0}=3$~Mm. 
The upper part of the signal propagates earlier than the rest of the pulse and strongly influences on phase-mixing.
On the other hand, a smaller arcade results in less diverged magnetic lines
and less $\grad c_{\rm A}(x,y)$,
that in consequence significantly reduces the fanning process of the wave signal.

Such a smaller arcade causes that a wave signal ranges all arcade region.
At $t=20$~s (panel $b$)
and later on, the outer edge of the arcade with its large 
length clearly reveals a phase difference of transversal wave velocity profiles 
as compared to the same in the core of the arcade. 
We observe several wave signal reflections from the transition region in Fig.~\ref{fig:6panels_3Mm}.
The violet-white patch in panels $d$ and~$e$ represents reflected Alfv\'en waves that propagate to the left-hand side.
When they reach the transition region near $x=3$~Mm,
they become reflected and propagate again into the right-hand side (panels $f$ and~$g$)
to suffer reflection once again 
near $x=16$~Mm,
which is clearly seen at $t=44$~s (panel $h$),
and go back into $x=3$~Mm (panels $i$ and~$j$).
It should be 
noted that the initial pulse-width
was assumed to be the same as 
in the case of $x_{\rm 0}=2$~Mm (Fig.~\ref{fig:6panels_2Mm}).
%
%
\subsubsection{Comparison}
%
%
In the presented vertical cross-sections of the Alfv\'en wave profiles,
we observe difference in wave propagation for large (Fig.~\ref{fig:cross-section}, top panel)
and small arcades (Fig.~{\ref{fig:cross-section}}, bottom panel).
The cross-section for the case of $x_{\rm 0}=2$~Mm ($x_{\rm 0}=3$~Mm)
shows the region near the apex of the arcade, $x=12$~Mm ($x=9$~Mm),
and three moments of time, when the wave signal passing the apex:
$t=22$~s ($t=20$~s), $t=72$~s ($t=40$~s) and $t=124$~s ($t=60$~s),
that are illustrated in the top (bottom) panel of Fig.~\ref{fig:cross-section}
by solid, dashed and dotted line, respectively.
The main difference between these two cross-sections
is a compact shape of the Alfv\'en wave in the short arcade during all three moments of time
in comparison to the large arcade.
This is a result of less diverged magnetic lines,
which spread the signal up to width equal $5$~Mm in the case of $x_{\rm 0}=2$~Mm,
while in the case of $x_{\rm 0}=3$~Mm a pulse width is about $2$~Mm.

We also find a difference in negative phase amplitudes.
The Alfv\'en wave cross-section in the large arcade (Fig.~{\ref{fig:cross-section}}, top panel)
already after one period exhibits minima with negative values of $V_{\rm z}$ accompanying the main wave signal.
At $t=72$~s (dashed line),
the first opposite-sign phase with amplitude $V_{\rm z}\simeq-2.2$~km~s$^{-1}$ at $y\simeq8.8$~Mm
is a part of the elongated main wave signal
still following into the transition region located near $x=2$~Mm to suffer reflection
(Fig.~\ref{fig:6panels_2Mm}, panel $g$, upper violet white patch),
while the second opposite-sign phase with amplitude $V_{\rm z}\simeq-0.8$~km~s$^{-1}$ at $y\simeq6.6$~Mm
is a signal that 
lags 
behind 
the main Alfv\'en signal, 
well seen in panel $g$ of Fig.~\ref{fig:6panels_2Mm} (lower violet patch).
In later moment of time, at $t=124$~s (dotted line) the negative value of $V_{\rm z}$
has a similar form, however the amplitudes are 
$V_{\rm z}\simeq-1.4$~km~s$^{-1}$ at $y\simeq7.6$~Mm
and $V_{\rm z}\simeq-0.8$~km~s$^{-1}$ at $y\simeq6.7$~Mm.

In the case of a short arcade (Fig.~{\ref{fig:cross-section}}, bottom panel),
for which divergence of magnetic lines and the gradient of $c_{\rm A}$ are smaller, 
the opposite-sign phases at the cross-section states only 
$0.9$\% of the main wave signal amplitude equal to $V_{\rm z}\simeq5.6$~km~s$^{-1}$ at $t=20$~s (solid line),
$7$\% of the amplitude equal to $V_{\rm z}\simeq3.8$~km~s$^{-1}$ at $t=40$~s (dashed line)
and $15.2$\% of the amplitude equal to $V_{\rm z}\simeq2.5$~km~s$^{-1}$ at $t=60$~s (dotted line). 

Figure~{\ref{fig:ts}} illustrates the time-signatures of $V_{\rm z}$ collected 
at points near the apex of each arcade. The time-signatures for cases of 
$x_{\rm 0}=2$~Mm, $x_{\rm 0}=2.5$~Mm and $x_{\rm 0}=3$~Mm are presented in top, middle and bottom panels, 
respectively.  
We set the detection point for the case of $x_{\rm 0}=2$~Mm to 
be at ($x=12$~Mm, $y=7$~Mm), for the case of $x_{\rm 0}=2.5$~Mm to 
be at ($x=11$~Mm, $y=5.1$~Mm) 
and for the case of $x_{\rm 0}=3$~Mm at 
($x=9$~Mm, $y=4.2$~Mm).  
Time-signatures for smaller arcades ($x_{\rm 0}=2.5$~Mm and $x_{\rm 0}=2$~Mm)
have a regular harmonic shape,
but the time-signature for the large arcade exhibits action on fanning process
that results in superimposition of incoming and reflected wave signals in $V_{\rm z}$
and finally in non-regular shape of $V_{\rm z}$ oscillations.
Note that the amplitude of the Alfv\'en waves decays 
with time and that $V_{\rm z}$ oscillates with its characteristic wave period 
equal to $P=50.7$~s for $x_{\rm 0}=2$~Mm, $P=27.2$~s for $x_{\rm 0}=2.5$~Mm and $P=20.4$~s for $x_{\rm 0}=3$~Mm.
These different wave periods result from the different arcade lengths
and different Alfv\'en speeds along the corresponding magnetic lines 
for $x_{\rm 0}=2$~Mm, $x_{\rm 0}=2.5$~Mm and $x_{\rm 0}=3$~Mm.
%

%
\begin{figure}
\centering{
           \includegraphics[width=7.2cm,angle=0]{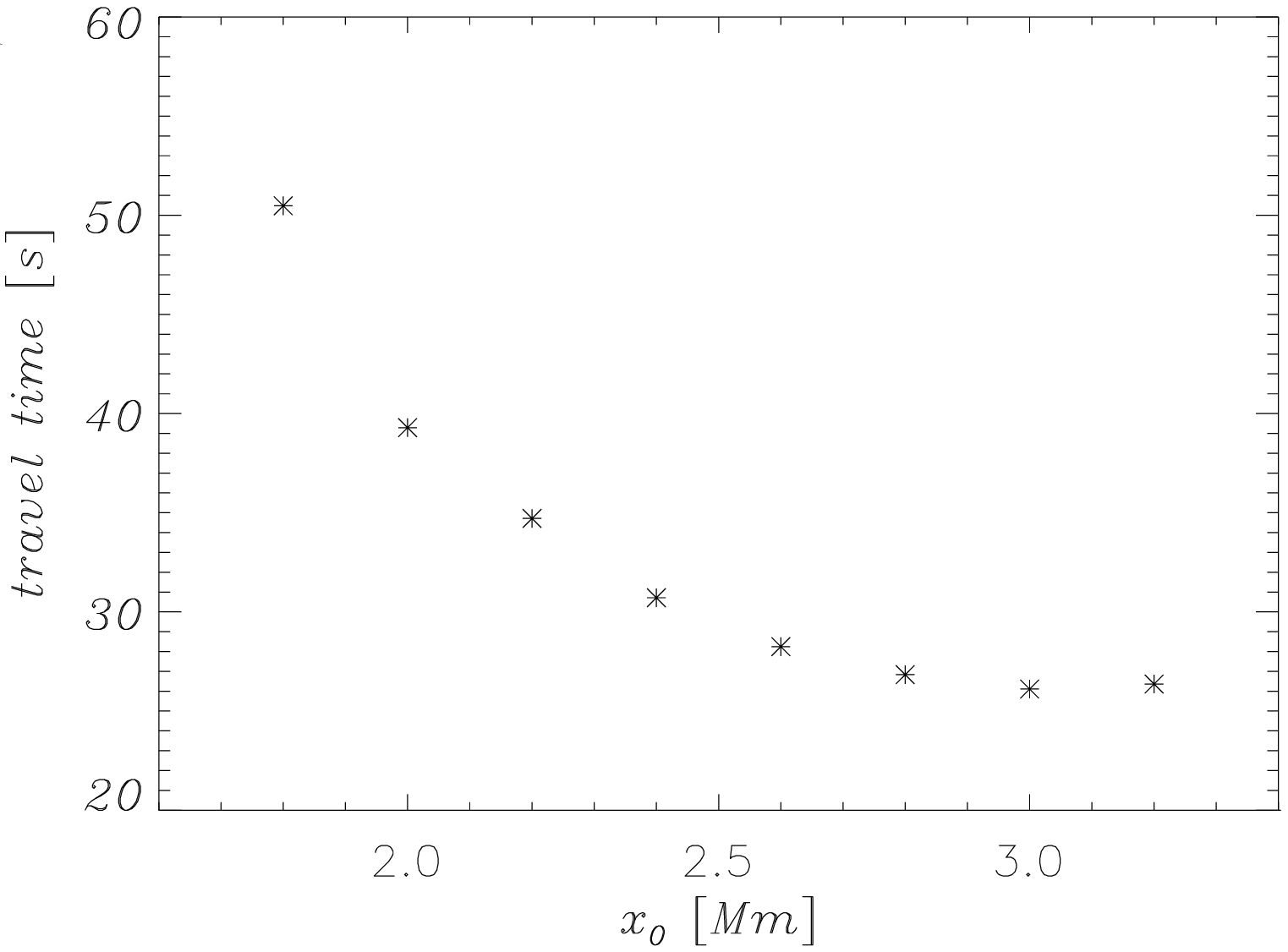}
          }
\caption{\small
         Travel time vs. the horizontal initial position, $x_{\rm 0}$,
	 along the magnetic field line crossing the point $(x_{\rm 0},1.75$~Mm$)$.
        }
\label{fig:tt}
\end{figure}
%

We calculate a travel time of the Alfv\'en waves, $t_{\rm t}$, for the first bounce
along few magnetic field lines,
which 
cross 
the given point $(x_{\rm 0},y_{\rm 0})$,
%
\beq
\label{eq:tt}
t_{\rm t} = \int_{0}^{l} \frac{d s}{c_{\rm A}(s)}\, .
\eeq
%
Here, a coordinate along the magnetic line has the value of $s=0$ at $(x_{\rm 0},y_{\rm 0})$,
while for $s=l$ the magnetic line ends at the transition region, $y=2.75$~Mm.
In Fig.~\ref{fig:tt} the travel time is displayed depending on
the initial point $x_{\rm 0}$.
Its value gradually 
decreases 
with $x_{\rm 0}$
from $t_{\rm t}\simeq39$~s at $x_{\rm 0}=2$~Mm, 
reaching $t_{\rm t}\simeq26$~s at $x_{\rm 0}=3$~Mm.
Because of almost linear decrease of the magnetic field length
from $l=23.4$~Mm at $x_{\rm 0}=2$~Mm up to $l=11.6$~Mm at $x_{\rm 0}=3$~Mm (not shown),
these results show that the time of the Alfv\'en wave traveling 
along longer magnetic field lines is much bigger, than for shorter magnetic lines. 
This effect is especially important in case of the large arcade,
where the travel time falls off rapidly, 
contributing to the wave signal profile fanning.

Our results show the complex picture of Alfv\'en waves caused by wave reflection 
from the transition region, and difficulties in direct studies of the interaction 
of counter-propagating waves in the asymmetric solar arcade. Because of these 
difficulties, we instead focus on evaluating a global attenuation time.
This allows us to investigate integrally the attenuation processes affecting 
the transverse waves, like interactions of the counter-propagating waves, 
partial wave reflection from the transition region, and the wave energy leakage
along a curved magnetic field lines of the arcade.

%
\begin{figure}
\centering{
           \includegraphics[width=7.6cm,angle=0]{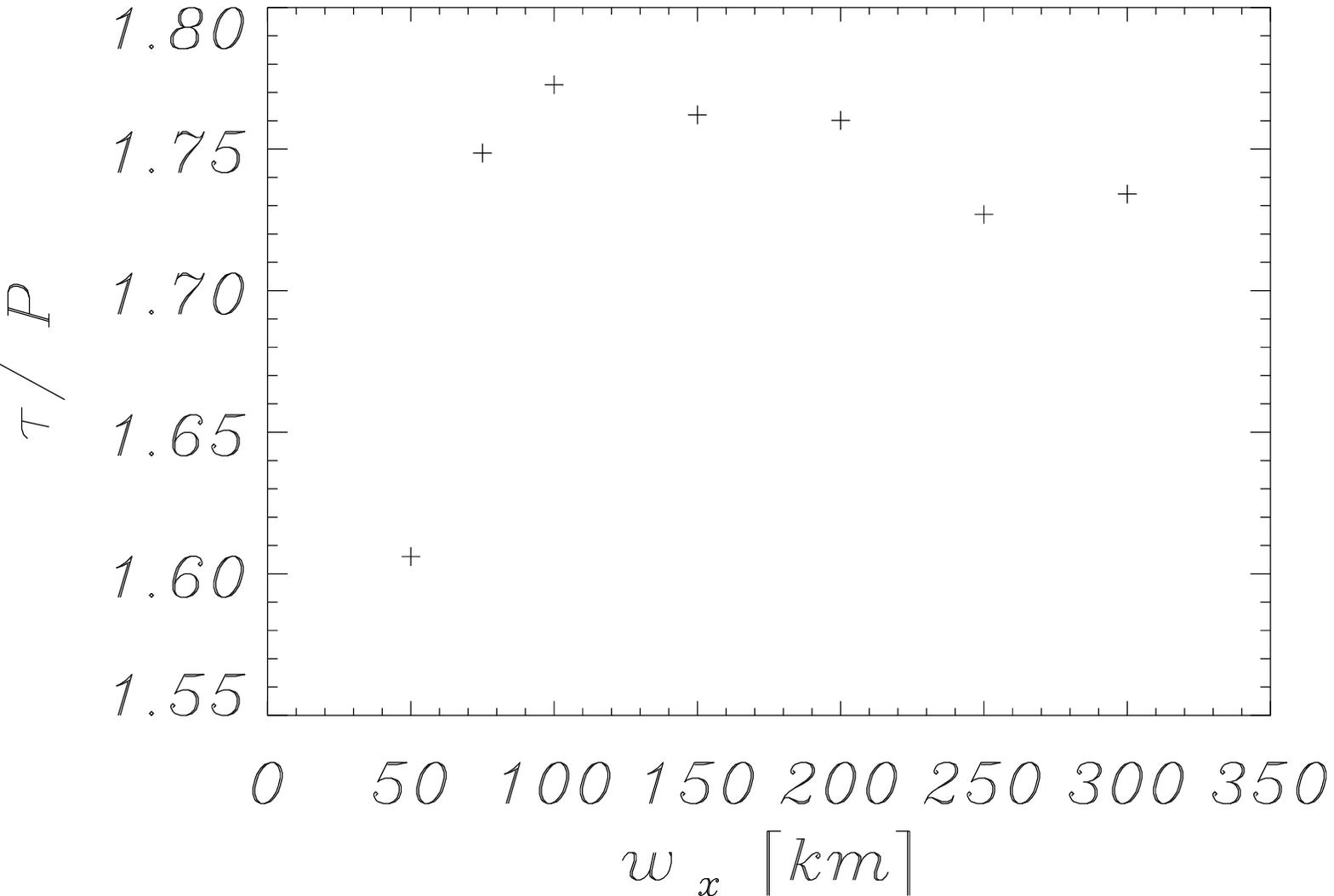}\vspace{0.2cm}\\
           \includegraphics[width=7.2cm,angle=0]{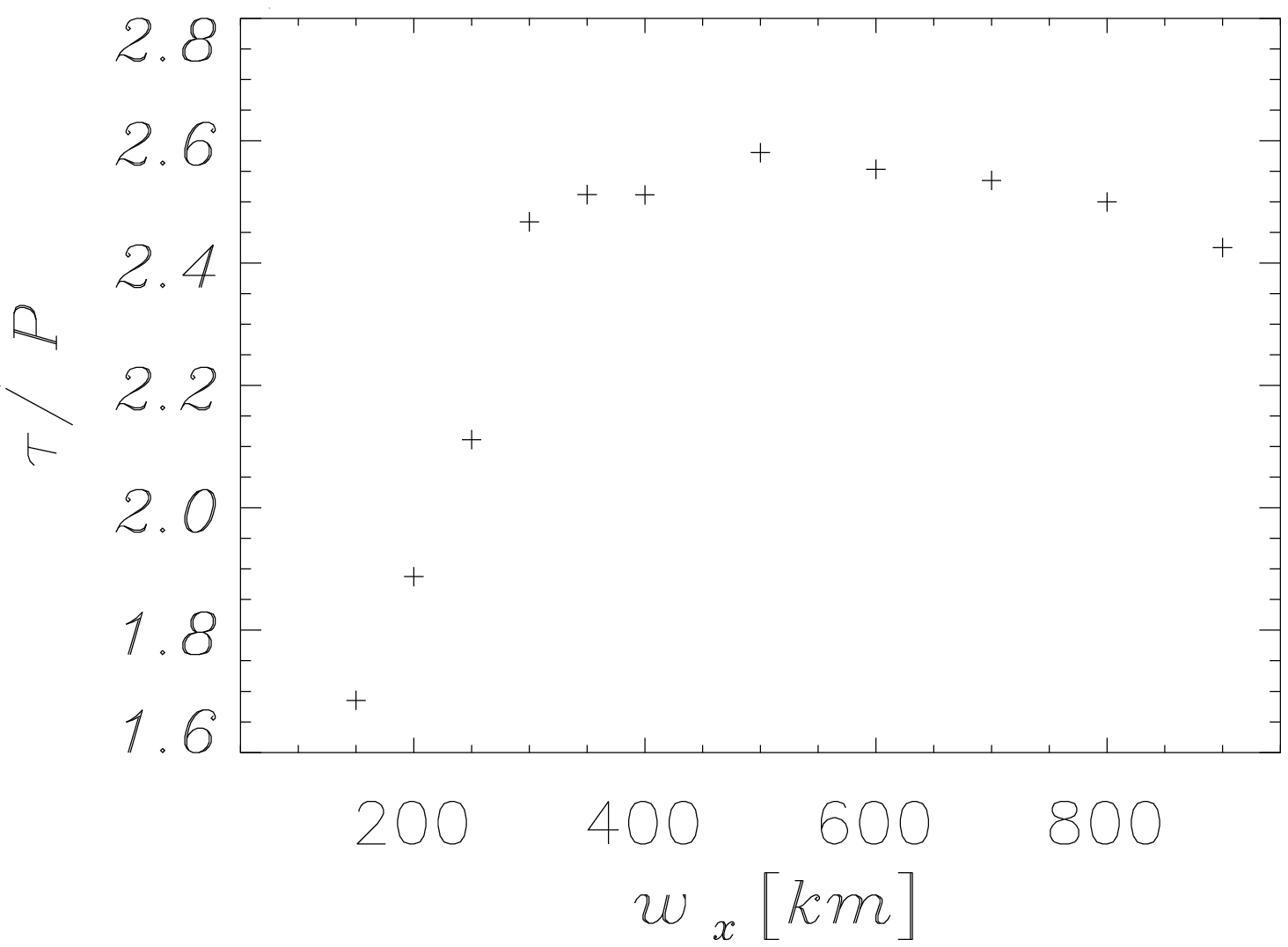}
          }
\caption{\small
         The ratio of the attenuation time over wave period, $\tau$/$P$, vs. pulse width $w_{\rm x}$
         for the case of $x_0 = 2$ Mm (top panel) and $x_0 = 3$ Mm (bottom panel).
        }
\label{fig:att}
\end{figure}
%

%
\begin{figure}
\centering{
           \includegraphics[width=8.4cm,angle=0]{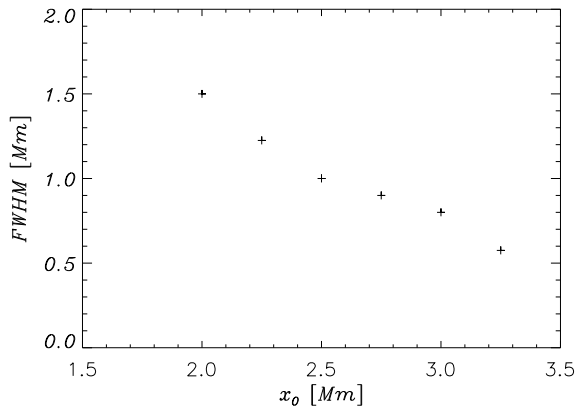}\vspace{0.2cm}
          }
\caption{\small
         The value of arcade full width at half maximum (FWHM) vs. $x_{\rm 0}$
         for the chosen $x=11$ Mm.
        }
\label{fig:width}
\end{figure}
%

%
\begin{figure}
\centering{
           \includegraphics[width=8.4cm,angle=0]{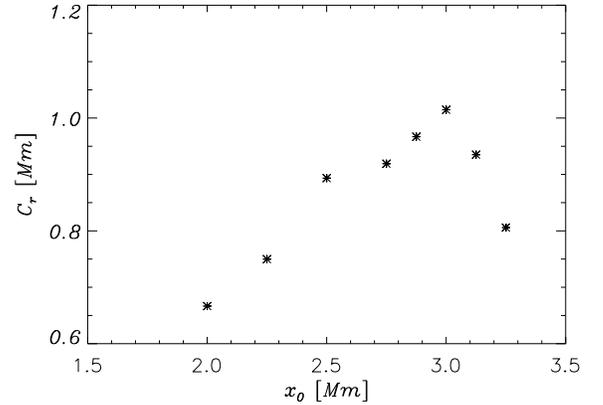}\vspace{0.2cm}
          }
\caption{\small
         The reflection coefficient, $C_{\rm r}$, vs. $x_{\rm 0}$.
        }
\label{fig:cr}
\end{figure}
%

%
\subsection{Different pulses width}
%

%
For a wider initial pulse, $w_{\rm x}$, the process of phase-mixing becomes stronger.
The reason 
is that both sides of such pulse experience a larger phase difference due to a larger 
difference in field lines curvatures on both magnetic interfaces
and pulse deformation process in the transition region as compared to a narrower pulse.
Hence, for the wider pulse more strongly Alfv\'en waves propagation is affected.

Basing on data signal of $V_{\rm z}$ collected at point ($x=12$~Mm, $y=7$~Mm)
for the case of longer arcade ($x_{\rm 0}=2$~Mm)
and ($x=9$~Mm, $y=4.2$~Mm) for the case of shorter arcade ($x_{\rm 0}=3$~Mm),
we evaluate a wave period $P$ using fast Fourier transform method and an attenuation time $\tau$.
The ratio of attenuation time $\tau$ to a wave period $P$ with respect to the width of the initial pulse
is presented in Fig.~\ref{fig:att}, which
shows that $\tau / P$ first increases for small values of $w_{\rm x}$ but 
then falls off slightly with $w_{\rm x}$.

When we look at the individual $\tau / P$ versus the pulse width profiles of 
the two types of arcades, they exhibit a similar trend that is some increment up 
to a certain pulse width and then decrement.  In the case of the considered geometry 
of more curved and larger magnetic arcade, the $\tau / P$ first increases up to $100$ km 
pulse width (Fig.~\ref{fig:att}, top panel), which means that up to this limit of the pulse width the attenuation 
of Alfv\'en waves decreases in the arcade.

Similarly in the case of the considered geometry of less curved and smaller 
magnetic arcade, the $\tau / P$ first increases up to $400$ km pulse width (Fig.~\ref{fig:att}, bottom panel),
which means that up to this limit of pulse width the attenuation of Alfv\'en waves decreases in 
the arcade. For $w_{\rm x}>400$~km, the $\tau / P$ decreases (Fig.~\ref{fig:att}, bottom panel),
which is the signature of the increment in the wave attenuation. In other words, for the Alfv\'en 
waves launched with the pulse widths larger than $400$~km, the generated long wavelengths  
do not fit within the considered loop geometry and the wave leakage again causes a 
wave attenuation (Gruszecki et al. 2007).

It is also found that the attenuation of the Alfv\'en waves is smaller (larger 
the attenuation time with respect to the wave period) in the more curved and 
larger magnetic arcades ($x_{\rm 0}=2$~Mm), while it is larger (smaller the 
attenuation time with respect to the wave period)  in the case of the less 
curved and comparatively smaller arcades ($x_{\rm 0}=3$~Mm). 
We found $\tau/P=1.75$~s for the core of $x_{\rm 0}=2$~Mm and $\tau/P=1.89$~s for $x_{\rm 0}=3$~Mm,
for $w_{\rm x}=0.2$~Mm.

The arcade full width at half maximum (FWHM) dependence on the position of initial pulse $x_{\rm 0}$,
and in consequence on a size of the arcade,
is shown in Fig.~\ref{fig:width}, which clearly
illustrates that a smaller arcade (bigger $x_{\rm 0}$) has a smaller width (FWHM);
a smaller arcade consists of weakly diverged magnetic field lines.

If one considers the phase-mixing to be the only candidate for the wave 
dissipation (e.g., Heyvaerts \& Priest 1983), then the strong phase-mixing 
in the more curved and larger arcades (cf., Fig.~\ref{fig:6panels_2Mm}) must 
cause a greater attenuation 
as compared to the less strong phase-mixing in the less curved and smaller 
arcades (cf., Fig.~\ref{fig:6panels_3Mm}).
Nevertheless, the simulation results demonstrate a more complicated scenario 
of the wave attenuation,
which simply implies that 
the phase-mixing may not be the only candidate responsible for the wave 
attenuation; actually, the magnetic field configuration and plasma properties of the arcade 
such as a spatial profile of the Alfv\'en speed, 
strength and dimension of the pulse, and the structure of the 
transition region, all play the vital role in this phenomenon.
%
\subsection{Partial reflection from the transition region}
%
Decay of the Alfv\'en wave amplitude results from the wave energy leakage 
caused by the curvature and divergence of magnetic field lines, inhomogeneous Alfv\'en speed $c_{\rm A}(x,y)$ 
and a partial 
penetration of Alfv\'en wave signal into the chromosphere (Gruszecki et al. 2007).
We clearly observe in our simulation that Alfv\'en waves
experience partial reflection
from the transition region and Alfv\'en waves signal penetrates
into solar atmospheric layers under the transition region (Figs.~\ref{fig:6panels_2Mm} 
and ~\ref{fig:reflection}, panels $d-h$). 
The amplitude of the waves penetrating into the lower region of the atmosphere 
drops from about $V_{\rm z}=1$~km s${}^{-1}$ at $t=48$~s just under the transition region
(Fig.~\ref{fig:reflection}, panel $d$), 
through $V_{\rm z}=0.5$~km s$^{-1}$ after about $16$~s (Fig.~\ref{fig:reflection}, panel $h$) 
and finally disappears. 
We can evaluate  the reflection coefficient,
%
\beq
\label{eq:cr}
C_{\rm r} = \frac{A_{\rm r}}{A_{\rm i}}\, ,
\eeq
%
where $A_{\rm i}$ ($A_{\rm r}$) is the amplitude of the incident (transmitted) waves.
Substituting $A_{\rm i}=6$~km~s${}^{-1}$
and $A_{\rm r}=4.01$~km~s${}^{-1}$
into the above formula we get $C_{\rm r}=0.668$,
which means that $66.8$\% of the waves amplitude became reflected
in the transition region and $33.2$\% was transmitted into lower atmospheric layers.
The reflection coefficient, $C_{\rm r}$, 
vs. initial pulse position $x_{\rm 0}$ for the first reflection from the transition region
is presented in Fig.~\ref{fig:cr}.
We expect that the amplitude of the wave signal reflected in the transition region
is smaller for Alfv\'en waves in larger inclined magnetic field lines,
which corresponds to larger values of $x_{\rm 0}$.

In the case of the largely curved 
arcades, the lateral leakage of the Alfv\'en wave energy may also be largely dominant 
as compared to the one associated with less curved arcades (Gruszecki et al. 2007).
The partial Alfv\'en wave reflection results from a steep gradient of 
Alfv\'en speed in the transition region because of a significant {mass} density drop 
in this region of the solar atmosphere.

{
It must be also noted that in case of linear Alfv\'en waves generated by small initial 
velocity pulses, $V_{\rm z}=3$~km~s$^{-1}$ with the maximum amplitude reaching only 
$V_{\rm z}\approx6$~km~s$^{-1}$, there are no associated density variations.}

%
\section{Summary and Conclusions}\label{sec:Summary}
%
To determine the role played by Alfv\'en waves in the coronal heating and solar 
wind acceleration, a number of authors investigated physical processes responsible 
for the excitation and attenuation of Alfv\'en waves in the solar atmosphere (e.g., 
Priest 1982; Ofman \& Davila 1995; {Ofman~2002; Miyagoshi et al.~2004;}
Dwivedi \& Srivastava 2006; {Ofman \&~Wang~2007;} Chmielewski et al.~2013,
and references therein).  In general, Alfv\'en waves are difficult to 
dissipate their energy and possible mechanisms involve the collisional dissipative 
agents, such as viscosity and resistivity, or non-classical plasma processes, such 
as mode-coupling and phase-mixing (cf., Heyvaerts \& Priest 1983; Nakariakov et al. 
1997; Zaqarashvili et al. 2006; Dwivedi \& Srivastava 2006, and references there).  
Overall, the collisional dissipative processes are found to be less important for 
the Alfv\'en wave dissipation than the non-classical plasma processes, especially 
phase-mixing (e.g., Nakariakov et al. 1997).  Extensive 
observational searches were performed to find signatures of the Alfv\'en waves 
dissipation in the solar corona (e.g., Banerjee et al. 1998; Harrison et al. 
2002; O'Shea et al. 2005; Bemporad et al. 2012, and references therein).  
However, as of today, there is no convincing observational evidence for the 
existence of Alfv\'en waves dissipation in the solar atmosphere.

In this paper, we simulated impulsively generated Alfv\'en waves in a 
stratified and magnetically confined solar arcade with the VAL-C temperature 
profile (Vernazza et al. 1981) as an initial realistic plasma condition in the 
curved magnetic field topology. Asymmetric solar magnetic arcades and the 
Alfv\'en waves propagation in these arcades were modeled by the time-dependent 
MHD equations that were solved numerically by the publicly available FLASH code 
(Lee \& Deane 2009).  {We analyzed the effects of changing the horizontal 
position of the initial pulse and its width on the Alfv\'en wave propagation 
in the asymmetric solar arcade}, the Alfv\'en wave deformation and phase-mixing 
resulting from inhomogeneous Alfv\'en wave velocity, and different lengths and 
divergence of magnetic field lines, and the partial wave reflection in the 
solar transition region.

We found that the more curved and larger arcade, then the stronger attenuation 
of the Alfv\'en waves.  Our results also demonstrated attenuation of Alfv\'en 
waves resulted from the curvature and divergence of magnetic field lines, 
inhomogeneous Alfv\'en wave velocity, partial reflection in the solar transition 
region, as well as the decrement of the attenuation time of Alfv\'en waves for 
wider initial pulses in the given magnetic configuration of various types of 
arcades.  {Moreover, our numerical simulations also showed that Alfv\'en 
waves, which are partially reflected in the solar transition region return 
to the solar chromosphere as slowly downward propagating Alfv\'en waves.}

The approach presented in this paper allowed us to investigate the effects 
caused by trains of Alfv\'en waves propagating in our arcade model as a 
result of the wave reflection in the solar transition region.  {Our 
results clearly show that the pulse deformation process in the transition 
region has a strong influence on the Alfv\'en waves propagation and on the
wave damping in the curved magnetic field of the solar arcade together with 
the asymmetric magnetic field configuration, the plasma properties of the 
arcade, the horizontal size of the pulse, and the structure of the solar 
transition region.}

{In the previous work, Nakariakov et al. (1999) investigated the
transverse oscillations of EUV loops as observed by the Transition 
Region and Coronal Explorer (TRACE), and demonstrated that the damping 
time was three times longer than the period of oscillations.  They 
considered kink oscillations of coronal loops and their damping caused 
by non-classical viscosity and resistivity.  Then, Ofman et al. (2002) 
showed that the dissipation of Alfv\'en waves due to chromospheric wave 
leakage was not sufficient to describe such fast damping of the observed 
transversal oscillations in the solar atmosphere.  Finally, Del Zanna et 
al. (2005) demonstrated that dissipation of Alfv\'en pulses within their 
coronal arcade led to a change in the local Alfv\'en speed at various 
heights, and depending upon various localized conditions that determine 
the observed damping of these transverse oscillations.} 

{In our work, the damping/attenuation time of the Alfv\'en waves are almost 
three times longer than that computed in both the longer and shorter arcades 
for a given pulse-width of 0.2 Mm.  It should be noted that this damping time
is due to the Alfv\'en wave dissipation caused by the phase mixing.  The ratio 
of attenuation time and wave period is almost the same to what is observed for 
the transverse wave dissipation in coronal loops.  An important point that must
be mentioned is that transverse kink waves with their radial velocity perturbations 
in the cylindrical thin tube geometry (Roberts 2000) are different than Alfv\'en 
waves studied in this paper, which are essentially azimuthal.  Nevertheless, there 
is a similar dissipation mechanism (e.g., phase-mixing) that may work for these 
two types of waves; its efficiency highly depends on the localized plasma, magnetic 
field configuration, nature of wave drivers, and Alfv\'en velocity, and others 
(see Ofman et al. 2002, 2007; Del Zanna et al. 2005).  Therefore, our parametric 
studies presented in this provide some likely physical scenario for dissipation 
of Alfv\'en waves through phase-mixing.}

Finally, we would like to point out that our approach is significantly different 
from that developed by Del Zanna et al. (2005), who considered a symmetric coronal 
arcade, which was embedded in the isothermal plasma with the constant (in time) 
temperature specified by the hyperbolic tangent profile, and designed it in such 
a way that it described short lived Alfv\'en waves in post flare settings. 
{Similarly, Miyagoshi et al. (2004) simulated Alfv\'en waves that were excited by 
perturbations imposed in the solar corona}.  By using our more realistic arcade 
model {with the asymmetric magnetic field configuration,} we were able to 
explore different physical aspects of the Alfv\'en wave propagation than those 
studied in the past.  Essentially, our model significantly generalizes the model 
originally considered by Del Zanna et al. (2005) {and by Miyagoshi et al.~(2004)}.
Nevertheless, further improvements of our model are possible and it is our hope 
that they would lead to even better understanding of the inherent physical 
complexity of phase-mixing process and its role in heating of the solar 
corona by Alfv\'en wave dissipation. 
%

%
\acknowledgments

%
%
We thank the referee for his/her valuable suggestions that improved our
manuscript considerably. 
P.Ch. expresses his cordial thanks to Piotr Konkol for his assistance in drawing some numerical data.
This work has been supported 
by NSF under the grant AGS 1246074 (K.M. \& Z.E.M.), and by the Alexander von Humboldt 
Foundation (Z.E.M.).
The software used in this work was in part developed by the 
DOE-supported ASC/Alliance Center for Astrophysical Thermonuclear Flashes at the 
University of Chicago.
%
%
%

%
%

%


\end{document}